\documentclass[12pt]{article}
\usepackage[utf8]{inputenc}

\usepackage{cite}
\usepackage[colorlinks=true,linkcolor=blue, linktoc=page, urlcolor=blue,citecolor=magenta]{hyperref}
\usepackage{amsmath}
\usepackage{amssymb}
\usepackage{float}
\usepackage{soul}

\usepackage{amsfonts}
\usepackage{dsfont}

\usepackage{graphicx}
\usepackage[dvipsnames]{xcolor}

\def\calo{{\cal O}}

\def\calr{{\cal R}}

\def\a{\alpha}
\def\b{\beta}
\def\g{\gamma}

\def\e{\epsilon}

\def\L{\Lambda}

\def\m{\mu}
\def\n{\nu}
\def\o{\omega}
\def\O{\Omega}

\def\r{\rho}

\def\rb{r_{\mathrm{b}}}
\def\rc{r_{\mathrm{c}}}
\def\rbz{r_{\mathrm{b}0}}
\def\rcz{r_{\mathrm{c}0}}

\newcommand{\os}[2]{{\overset{\,\scalebox{0.5}{($#2$)}}{#1}}{}}

\begin{document}

	\begin{titlepage}
		\vskip 1.8 cm
		
		\begin{center}{\huge \bf Dynamical de Sitter black holes in a quasi-stationary expansion}\\
			
		\end{center}
		\vskip 1cm
		
		\begin{center}{\large {{\bf Aaron Beyen$^1$, Efe Hamamc\i$^2$,\\ \vspace{0.1cm}  Kasper Meerts$^1$ and Dieter Van den Bleeken$^{1,2\,*}$}}}\end{center}
		
		\vskip 0.6cm
		
		\begin{center}
			1) Institute for Theoretical Physics, KU Leuven\\
			3001 Leuven, Belgium
			
			\vskip 0.3cm
			
			2) Physics Department, Boğaziçi University\\
			34342 Bebek / Istanbul, Turkey
			
			\vskip 1cm
			
			\texttt{aaron.beyen, kasper.meerts @kuleuven.be\\ efe.hamamci@boun.edu.tr, dieter.vdbl@gmail.com}
		\end{center}
		\vskip 0.5cm \centerline{\bf Abstract} \vskip 0.2cm \noindent 
			We revisit and improve the analytic study \cite{Gregory:2018ghc} of spherically symmetric but dynamical black holes in Einstein's gravity coupled to a real scalar field. We introduce a series expansion in a small parameter $\epsilon$ that implements slow time dependence.  At the leading order, the generic solution is a quasi-stationary Schwarzschild-de Sitter (SdS) metric, i.e. one where time-dependence enters only through the mass and cosmological constant parameters of SdS. The two coupled ODEs describing the leading order time dependence are solved up to quadrature for an arbitrary scalar potential. Higher order corrections can be consistently computed, as we show by explicitly solving the Einstein equations at the next to leading order as well. We comment on how the quasi-stationary expansion we introduce here is equivalent to the non-relativistic $1/c$ expansion.
			
		\vspace{3.2cm}
		
		\noindent\footnotesize{*Currently at Department of Meteorological and Climate Research,
			Royal Meteorological Institute, 1180 Uccle, Belgium}
	\end{titlepage}

\thispagestyle{empty}
\addtocounter{page}{-1}
		
		\tableofcontents

\newpage
\section{Introduction}\label{intro}
Physical black holes are quite dynamic, they interact with surrounding matter, collide with each other and probably continuously emit Hawking radiation. In addition, the surrounding space-time undergoes cosmological evolution. When it comes to analytic black hole solutions to Einstein's equations, these are however mainly restricted to stationary cases, with only a few rather simple exceptions such as the Vaidya \cite{Vaidya1951} and McVittie \cite{mcvittie1933mass} metrics.
\\
After some earlier work \cite{Terashima2000, Frolov2002,Harada2004,Sultana2005,UrenaLopez2011,Babichev2012,Davis2016} and building upon \cite{Chadburn2013,Gregory2017}, important progress was made in \cite{Gregory:2018ghc} towards the dynamics of spherically symmetric black holes in the regime of slow, quasi-stationary, time evolution. They were interested in the role of black holes in the inflationary phase of the universe and this motivated them to consider the black hole evolution under slow roll conditions similar to those of textbook inflation. There are other good reasons why the regime of slow time evolution is particularly interesting. Our work is mainly motivated by an attempt to generalize the usual black hole thermodynamics \cite{ black_hole_physics_book, Lambda, Dolan_2013, Kastor_2009} to a non-equilibrium setting, where quasi-stationary evolution and the corrections to it play a fundamental role. This new approach to black hole thermodynamics is essential because, for example, Schwarzschild black holes have a negative heat capacity \cite{Bhattacharya_2017}, which is incompatible with equilibrium thermodynamics. Following \cite{Maes_2019,nonequilibrium_steady_states}, it turns out that the heat capacity can still be calculated through a non-equilibrium method by using quasi-static evolution as worked out in this paper. This non-equilibrium derivation of the heat capacity will be discussed in the forthcoming paper \cite{Beyen23}.
\\Another somewhat unrelated, motivation comes out of the study of the strong field, non-relativistic limit of general relativity. This regime has mainly been investigated through an expansion in the inverse speed of light, the so-called $1/c$ expansion. For this setup, it has recently been shown in \cite{Elbistan2022} that it can equivalently be recast as a perturbative approximation of slow time dependence.
\\
\\
In this paper we revisit the analysis of \cite{Gregory:2018ghc} concerning spherically symmetric solutions to Einstein gravity minimally coupled to a single real scalar field $\Phi$, i.e. the theory described by the Lagrangian
\begin{equation}
L=\sqrt{-g}\left(\, R-\frac{1}{2}\partial_\mu\Phi \partial^\mu\Phi-V(\Phi)\right)\label{modellag}
\end{equation}
Closely following \cite{Gregory:2018ghc}, we work in coordinates where $\Phi$ is a function of a preferred time $t$ only and analyse the $t$-dependence of the scalar field and the metric perturbatively. 
\\ However, we improve on \cite{Gregory:2018ghc} in several ways:
\begin{enumerate}
	\item 
We introduce, in section \ref{setup}, a single small parameter $\epsilon$ through a redefinition of time as $\tau=\epsilon t$ together with a rescaling of the scalar field $\varphi(\tau)=\sqrt{\epsilon}\Phi(t)$. That way the slow time expansion takes the precise form of a power series expansion in $\epsilon$ which allows for a solution of the Einstein equations order by order. Compared to \cite{Gregory:2018ghc}, the introduction of $\epsilon$ allows for a clear identification of the order of the various terms that appear and thus for a more systematic formulation of the expansion.
\item We show that, to have a solution that can be extended beyond the leading order, one needs to generalize the metric ansatz used in \cite{Gregory:2018ghc}. As discussed in section \ref{section 3.1} and appendix \ref{notimeap}, a more general timelike warpfactor is required from order $\calo(\epsilon^1)$ onwards.

\item Already at leading order $\mathcal{O}(\epsilon^0)$, our solution differs slightly from that of \cite{Gregory:2018ghc} in that ours is fully quasi-stationary, while the metric of \cite{Gregory:2018ghc} is only partially quasi-stationary at this order. For the precise details see subsection \ref{comparison with 1}, but let us elaborate here a little. Under the assumptions of vanishing time derivatives at leading order  $\mathcal{O}(\epsilon^0)$, the solution to the Einstein equations for the spherically symmetric ansatz we consider has two integration constants: $(m,\Lambda)$. They represent the mass and cosmological constant of the static Schwarzschild de Sitter (SdS) spacetime respectively. In our setup, the Einstein equations imply, at leading order, a replacement $(m,\Lambda)\rightarrow (m(\tau),\Lambda(\tau))$ in {\it all} metric components of SdS. This makes our solution at leading order a quasi-stationary version of SdS, i.e. one where time evolution takes place as a motion in the parameter space of static solutions. Note that the functions $(m(\tau),\Lambda(\tau))$ are determined by two ODEs implied by the Einstein equations.  The leading order metric of \cite{Gregory:2018ghc} is very similar, but it is only {\it partially } quasi-stationary, since the replacement $(m(\tau),\Lambda(\tau))$ takes place only in {\it part of} the metric components, while in some components $(m,\Lambda)$ remain constant \footnote{More precisely, in \cite{Gregory:2018ghc} the replacement $(m,\Lambda)\rightarrow (m(\tau),\Lambda(\tau))$, which is equivalent to a replacement $(\rb,\rc)\rightarrow (\rb(\tau),\rc(\tau))$, takes place in $f$, i.e. $f\rightarrow f_{QS}$ but not in $\eta$, see (38, 39) in \cite{Gregory:2018ghc}. We discuss this in more detail in subsection \ref{comparison with 1}. \color{black}}. 

\item Where we implement the same physical boundary conditions on the horizons as \cite{Gregory:2018ghc}, we do however consider them order by order in our expansion parameter $\epsilon$. This procedure guarantees one finds a set of ODEs at each order of the expansion that determines the time evolution uniquely and which allows for an order-by-order solution, see sections \ref{LO} and \ref{NLO}. \\
\\

\item Finally, in section \ref{NLO}, we integrate the next-to-leading order, i.e. $\mathcal{O}(\epsilon^1$), Einstein equations explicitly, making the precise radial dependence of the metric manifest, where in \cite{Gregory:2018ghc} this dependence was left implicit in terms of some radial integrals. After the radial integration at this order, the two remaining degrees of freedom in the metric are determined by the boundary conditions through a simple ODE and an algebraic relation. It turns out that the ODE can be solved by direct integration, but we have not been able to simplify the integrand to a form that is worth reproducing here, let alone integrate explicitly.

\end{enumerate} 

\color{black}
As we already briefly mentioned, the solution we will present is also of relevance to the nonrelativistic $1/c$ expansion of general relativity, which has recently been quite actively researched, see \cite{Hartong2022} for a review. We'll show how, upon identification of the slow time parameter $\epsilon$ with $1/c$, the quasi-stationary expansion as introduced in this paper is indeed equivalent to the $1/c$ expansion of \cite{Elbistan2022}. 
\\There are two important contributions this paper makes to the study of the $1/c$ expansion. First of all, non-trivial explicit solutions to the $1/c$ expansion that fall outside of the weak field post-Newtonian regime have so far been restricted only to expansions of known exact relativistic solutions of GR, see e.g. \cite{Bleeken2019, Ergen2020, Hansen2020}. The solution presented here is, as far as we are aware, thus the first example of a solution to a few orders of the $1/c$ expansion for which no exact all order, or equivalently full relativistic GR, solution is known. Secondly, the $1/c$ expansion consists of a set of iteratively solvable PDEs in terms of the spatial derivatives of the fields, with time derivatives of lower order fields only appearing as source terms in the equations for the higher order fields (see e.g. \cite{Elbistan2022} for a detailed discussion). As such it remained unclear if and how the time dependence of the fields is constrained. Here this issue is resolved through imposing boundary conditions for the fields, following \cite{Gregory:2018ghc}, which provide ODEs that determine the time dependence, and hence dynamics, of the fields. Our work could thus be a concrete example and starting point towards a better understanding of how the dynamics should be implemented in the $1/c$ expansion. We believe it should be possible to formulate generic types of boundary conditions that would do this for a general metric.
\\
\\
This paper is organized as follows. We begin in section \ref{SdSsection} by reviewing the Schwarzschild-de Sitter geometry, which solves the Einstein equations in the presence of a positive cosmological constant. It is a two-parameter family of solutions and we spend some time recalling different choices of parameters and their relations as these will play some role in the later sections. Section \ref{sphericalsymsec} discusses 
the search for dynamical spherically symmetric solutions to Einstein gravity minimally coupled to a real, time-dependent scalar field. To this end, we choose coordinates adapted to the problem at hand in which the most general spherically symmetric metric ansatz is parametrised. We then show how the Euler-Lagrange equations reduce to two coupled PDEs. In addition, we discuss the remaining coordinate freedom, as well as introduce and motivate the boundary conditions \cite{Gregory:2018ghc} that will be imposed. 
\\The new and most important results of this paper find themselves in sections \ref{quasi-stationary expansion} and \ref{solsec}. More specifically, section \ref{quasi-stationary expansion} introduces the small parameter $\epsilon$ and explains in detail how the equations of motion 
can be solved through a series expansion in $\epsilon$. We also show how this expansion in $\epsilon$ is equivalent to the $1/c$ expansion in section \ref{1csec}. In section \ref{solsec}, we then explicitly solve the problem for the first two orders in the expansion in $\epsilon$. There one sees how at leading order the solution is a quasi-stationary SdS metric with the time dependence determined through the boundary conditions and it is where the main differences and improvements concerning \cite{Gregory:2018ghc} can be found. 
\\Finally, there are three appendices to the paper. In appendix \ref{notimeap}, we show how the Einstein equations can be solved exactly when one restricts the metric ansatz to the form used in \cite{Gregory:2018ghc} and {restricts to a purely scalar field energy-momentum tensor. The key point is that there is {\it no} dynamic black hole among these solutions. This small result is the main motivation to consider the most general spherically symmetric ansatz in section \ref{sphericalsymsec}. Appendix \ref{bcapp}} mentions some technical details used in the motivation of our boundary conditions and lastly, appendix \ref{toyap} illustrates some of the features of the quasi-stationary expansion in the simple toy model of a damped oscillator.

\section{The Schwarzschild-de Sitter black hole}\label{SdSsection}
Before we start our study of time-dependent spherically symmetric black holes coupled to a scalar field, it will be useful to shortly review some key features of the well-known spherically symmetric and static Schwarzschild de Sitter (SdS) solution. It is a solution to the theory \eqref{modellag} where the scalar field is a constant $\Phi(x^\m)=\Phi_\star$, with $\Phi_\star$ an extremum of the potential $V$ that further determines a cosmological constant
\begin{equation}
\Lambda=\ell^{-2}=\frac{V(\Phi_\star)}{2}\,.\label{elldef}
\end{equation}
In static coordinates the SdS metric takes the form
\begin{equation}
ds^2_{\mathrm{SdS}}=-F dt_\mathrm{s}^2+F^{-1}dr^2+r^2 d\Omega^2 \, ,
 \label{SdSsol}
\end{equation}
where 
\begin{equation}
F=1-\frac{2m}{r}- \frac{\Lambda r^2}{3}\, .\label{fdef}
\end{equation}
The parameter $m$ has the physical interpretation of the mass of the black hole. When $\Lambda=0$, $m$ coincides with the usual mass of the Schwarzschild black hole and more generally speaking $m$ is the conserved charge related to the Killing vector $\partial_{t_\mathrm{s}}$.
We will restrict attention to the case where both $m$ and $\Lambda$ are positive. \\In this paper, rather than using the parameters $(m,\Lambda)$ to label the family of SdS geometries, the equivalent pair $(\alpha,\ell)$ is often more convenient. Here $\ell$ is the cosmological length scale \eqref{elldef} and $\alpha$ is an angle defined via
\begin{equation*}
m=\frac{\ell\cos 3\alpha}{3}\,,\qquad 0\leq \a\leq \frac{\pi}{6}\,.
\end{equation*}
One can verify that \eqref{fdef} can be rewritten as
\begin{equation}\label{F wth roots}
F=-\frac{1}{3\ell^2r}(r-\rc)(r-\rb)(r+\rc+\rb)\,,
\end{equation}
where
\begin{eqnarray}
\rb&=&2\ell\cos(\a+\frac{\pi}{3})=\ell(\cos\a-\sqrt{3}\sin\a)\,, \label{adef}\\ \rc&=&2\ell\cos(\a-\frac{\pi}{3})=\ell(\cos\a+\sqrt{3}\sin\a)\,. \label{rcdef}
\end{eqnarray}
Remark from \eqref{F wth roots} that $0\leq r_\mathrm{b}\leq r_\mathrm{c}$ are the two positive roots of $F$ which determine the radial position of a black hole and cosmological horizon respectively. The choice $\alpha=\frac{\pi}{6}$ corresponds to the de Sitter solution, i.e. dS$_4$, while $\alpha=0$ represents the extremal case where $\rb=\rc$. Note that in the limit $\alpha\rightarrow 0$ the physical distance between the horizons does not go to zero, instead, the space-time metric between the horizons becomes that of dS$_2\times$S$^2$, see e.g. \cite{Anninos2012}.
\\
One can also use the pair $(\rb,\rc)$ to parametrize the family of solutions, in terms of these the mass and cosmological length scale take the form
\begin{equation*}
\ell^2=\frac{1}{3}\left(\rb^2+\rc^2+\rb \rc\right)\, ,\qquad  m=\frac{1}{2}\frac{\rc \rb (\rc+\rb)}{(\rb^2+\rc^2+\rb \rc)}\, .
\end{equation*}
Yet another parametrization of the family of SdS solutions is in terms of the pair $(M,\alpha)$ which is related to the $(m,\Lambda)$ parametrization as
\begin{equation}\label{r alpha l cos}
\Lambda=\frac{\cos^2(3\alpha)}{24\sqrt{3}M^2\sin^2\!\alpha\sin 2\alpha}\, ,\qquad m=\frac{2\sqrt{2}M}{3^{1/4}}\sin\a\sqrt{\sin 2\a}\, .
\end{equation}
Although this $(M,\alpha)$ parametrization appears ad hoc and certainly not very elegant, we will see in section \ref{LO} that it is crucial when studying slow time evolution, since under dynamic evolution $M$ remains constant and time dependence enters only through $\alpha$. The normalization of $M$ is chosen such that as $\alpha\rightarrow \frac{\pi}{6}$ and $M$ constant one finds $\Lambda\rightarrow 0$ and $m\rightarrow M$.  I.e. in the $(M,\alpha)$ parametrization the choice $(M,\frac{\pi}{6})$ corresponds to the standard asymptotically flat Schwarzschild solution of mass $m=M$.
\\
\\
The time coordinate $t_\mathrm{s}$, the 'static time', appearing in \eqref{SdSsol} is special in that $\partial_{t_\mathrm{s}}$ is a Killing vector and because \eqref{SdSsol} is time reversal invariant with respect to it, i.e. $t_\mathrm{s}\mapsto -t_\mathrm{s}$ is an isometry. But in the analysis of time-dependent generalizations of the SdS solution, it turns out useful and efficient to work with a new time coordinate $t$ chosen such that the scalar field is a function of $t$ only. This coordinate $t$ does not directly reduce to $t_\mathrm{s}$ in the special case of static solutions, rather it is related to $t_\mathrm{s}$ by the coordinate transformation \cite{Gregory:2018ghc}
\begin{eqnarray}
t&=&t_\mathrm{s}+\frac{1}{2\kappa_\mathrm{b}}\log \left(\frac{r}{\rb}-1\right)
-\frac{1}{2\kappa_\mathrm{c}}\log \left(1-\frac{r}{\rc}\right)\nonumber\\
&&-\frac{1}{4\kappa_\mathrm{b}\kappa_\mathrm{c}}\frac{\rc-\rb}{\rb\rc}\log\left(1+\frac{r}{\rb+\rc}\right)-\frac{\rc\rb}{\rc-\rb}\log\frac{r}{\rb+\rc}\label{hdef} \\
&=&  t_\mathrm{s} + h(r) \, ,\nonumber
\end{eqnarray}
where the $\kappa$'s are the surface gravities at the respective horizons
\begin{eqnarray*}
\kappa_\mathrm{b}&=&\frac{1}{2}|F'(\rb)|=\frac{(\rc-\rb)(2\rb+\rc)}{2\rb (\rb^2+\rc^2+\rb\rc)}\, ,\quad 
\kappa_\mathrm{c} = \frac{1}{2}|F'(\rc)|=\frac{(\rc-\rb)(2\rc+\rb)}{2\rc (\rb^2+\rc^2+\rb\rc)}\, .
\end{eqnarray*}

With respect to the time coordinate $t$, the metric \eqref{SdSsol} then takes the form
\begin{equation}
ds^2_{\mathrm{SdS}}=-F dt^2+2 H dt dr+\frac{1-H^2}{F}dr^2+r^2 d\Omega_2^2\, ,\label{stationSdS}
\end{equation}
where
\begin{equation}
H=\frac{\beta}{r^2}-\gamma r\, ,\qquad \beta=\frac{\rb^2\rc^2 (\rb+\rc)}{\rc^3-\rb^2}\, ,\qquad \gamma=\frac{\rb^2+\rc^2}{\rc^3-\rb^3}\, .\label{Hform}
\end{equation}
It is important to point out that $\partial_t=\partial_{t_\mathrm{s}}$, i.e. also $t$ is a choice of time adapted to the Killing vector. What distinguishes $t$ among such adapted time coordinates is that via \eqref{Hform} it guarantees that
\begin{equation}\label{boundary H}
H(\rb)=-H(\rc)=1\, .
\end{equation}
This condition ensures that $\partial_r$ is null on both horizons, something we'll further illustrate below. The property \eqref{boundary H} will be imposed as a boundary condition on the dynamical solutions, see section \ref{bcsec}, guaranteeing that the scalar field will be purely ingoing on both horizons.

\begin{figure}[H]
	\begin{center}
	\includegraphics[scale=0.5]{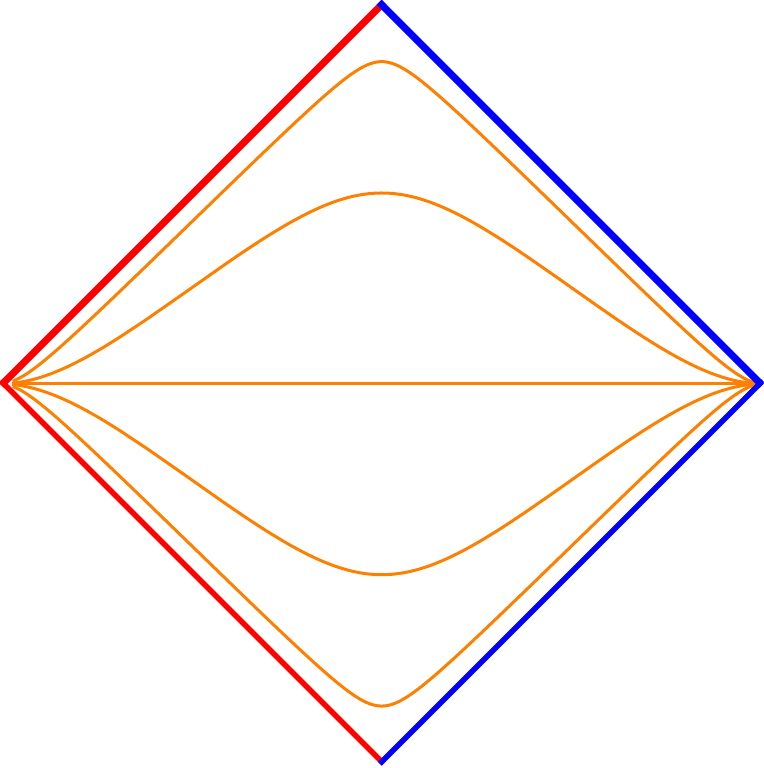}\qquad \includegraphics[scale=0.5]{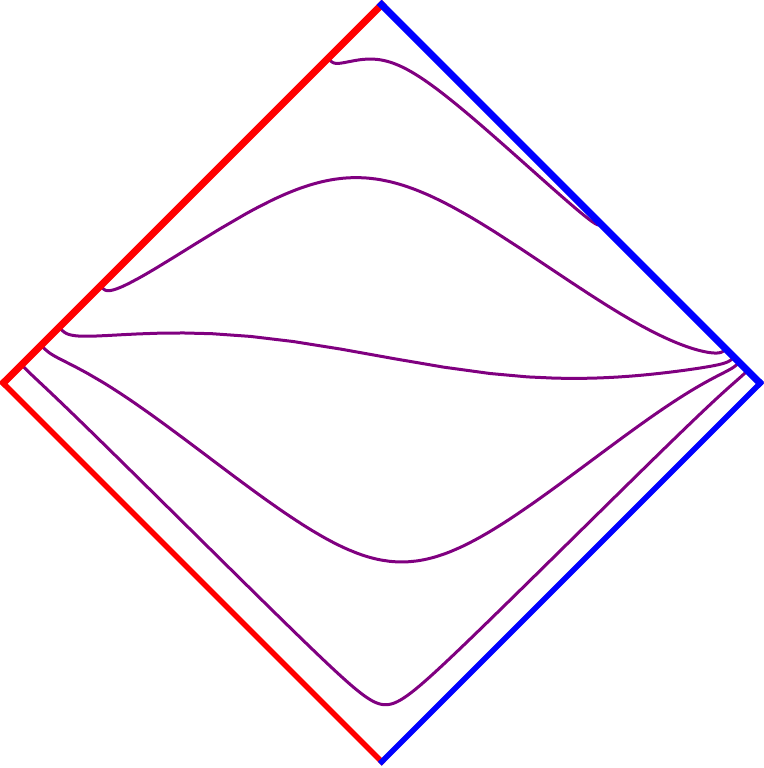}
	\end{center}
	\caption{Penrose diagram for the region $\rb\leq r\leq \rc$ of the SdS spacetime, the black hole horizon is in red, the cosmological horizon in blue. On the left constant $t_\mathrm{s}$ slices are shown in orange and on the right constant $t$ slices in purple.}
	\label{Penrose}
\end{figure}

\noindent The peculiarities of the time $t$ are most easily illustrated in the Penrose diagram of this spacetime. For simplicity, we only consider the Penrose diagram for the region between the horizons; for a more general discussion see e.g. \cite{kroon2023conformal}. One starts by introducing the tortoise coordinate
\begin{align}
	r_*&=\int ^r \frac{dr'}{F} = \frac{(r_\mathrm{b}^2+r_\mathrm{b}r_\mathrm{c}+r_\mathrm{c}^2)}{(r_\mathrm{c}-r_\mathrm{b})(2r_\mathrm{b}+r_\mathrm{c})(2r_\mathrm{c}+r_\mathrm{b})} \times\label{tortoise} \\
	& \Big(r_\mathrm{b}(r_\mathrm{b}+2 r_\mathrm{c})\log(r-r_\mathrm{b})-r_\mathrm{c}(r_\mathrm{c}+2 r_\mathrm{b})\log(r_\mathrm{c}-r)+(r_\mathrm{c}^2-r_\mathrm{b}^2)\log(r+r_\mathrm{b}+r_\mathrm{c})\Big)\, , \nonumber \label{tortoise}
\end{align}
and the associated null coordinates with a finite range
\begin{equation}
	U=\arctan(t_\mathrm{s}-r_*)\qquad V=\arctan(t_\mathrm{s}+r_*)\, ,\label{UV}
\end{equation}
so that the metric \eqref{SdSsol} takes the form
\begin{equation}
	ds^2=-\frac{F}{\cos U\,\cos V}dU dV\, .
\end{equation}
Since $r_*\rightarrow -\infty$ when $r\rightarrow r_\mathrm{b}$, one sees that $U=\frac{\pi}{2}$ and $V=-\frac{\pi}{2}$ correspond to the black hole horizon, while from $r\rightarrow r_\mathrm{c}$ corresponding to  $r_*\rightarrow \infty$, on sees that $U=-\frac{\pi}{2}$ and $V=\frac{\pi}{2}$ correspond to the cosmological horizon. Using the formulae (\ref{tortoise}, \ref{UV}), one can then plot the lines of constant $t_\mathrm{s}$ in the Penrose diagram, see the left diagram in Figure \ref{Penrose}. Additionally combining \eqref{hdef} with (\ref{tortoise}, \ref{UV}) allows one to plot the lines of constant $t$ as well, which is done in the right diagram in Figure \ref{Penrose}. One sees there explicitly that these lines of constant $t$ end perpendicularly on both the future black hole and cosmic horizon. This condition is equivalent to $\partial_r$ -- which in coordinates $(t,r)$ is tangent to the purple lines -- being null on these horizons.

\section{Spherically symmetric dynamics}\label{sphericalsymsec}
 \subsection{Einstein equations}\label{section 3.1}
The Euler-Lagrange equations of the Lagrangian \eqref{modellag} are
\begin{equation}
G^{\m\n}=\frac{1}{2}\left(\partial^\mu\Phi\partial^\nu\Phi-g^{\m\n}\left(\frac{1}{2}\partial_\m \Phi \partial^\mu \Phi + V(\Phi)\right)\right)\, ,\qquad \nabla_\mu \partial^\m \Phi - V'(\Phi)=0\, .\label{eoms}
\end{equation}
where $G^{\m\n}=R^{\m\n}-\frac{1}{2}g^{\m\n}R$ is the Einstein tensor. We will be interested in spherically symmetric solutions to this model. The most general spherically symmetric metric can be put into the form
\begin{equation}
ds^2=-{F} A^2dt^2+2HA  dt dr+\frac{1-H^2}{F} dr^2+r^2 d\Omega^2\, ,\label{metricform}
\end{equation}
where $F=F(t,r), H=H(t,r), A=A(t,r)$ are arbitrary functions of $t$ and $r$. The time coordinate $t$ is chosen\footnote{{Note that the choice \eqref{scalarform} can generically only be made locally in a patch and will break down where $\partial_t\Phi$ vanishes. We discuss this point in more detail at the end of subsection \ref{behaviour of the solutions}.}} such that the scalar field is a function of $t$ only, i.e. $\partial_r\Phi=0$ or:
\begin{equation}
\Phi=\Phi(t)\,.\label{scalarform}
\end{equation}
Remark that the choice \eqref{scalarform} prevents removing the $dtdr$ term in the metric. Of course, one could alternatively remove the $dtdr$ term by redefining $t$, but then $\Phi$ would, in general, be a function of both $t$ and $r$.

 Let us point out that the ansatz used in \cite{Gregory:2018ghc} corresponds to \eqref{metricform} with the additional assumption $A=1$. {We relax this condition as it obstructs the existence of non-trivial time-dependent black hole solutions beyond the quasi-stationary leading order $\mathcal{O}(\epsilon^0)$. Indeed, in appendix \ref{notimeap}, we show how the Euler-Lagrange equations with a scalar field \eqref{eoms} can be integrated with full generality when $A=1$ and that there are no time-dependent black hole solutions in this class. Furthermore in section \ref{NLO}, one can see explicitly that corrections to the quasi-stationary limit source the function $A$ and force it to differ from $1$.} 
\\
\\
The Euler-Lagrange equations \eqref{eoms} when evaluated on (\ref{metricform}, \ref{scalarform}) are equivalent\footnote{{More precisely, the metric ansatz \eqref{metricform} leads to the equalities
\begin{align*}
	\frac{r}{F}\Big(AH(E^{r}{}_r - E^{t}{}_t )- \frac{1-H^2}{F}E^{r}{}_t\Big)&=2H\partial_rA - \partial_t\left(\frac{1-H^2}{F}\right)\\
	-r^2 E^{r}{}_t&=\partial_t(rF) + \frac{r^2 H\dot\Phi^2}{2A}\\
	r^2 E^{t}{}_t&=\partial_r (rF) -1 + \frac{V}{2}r^2 + \frac{(1-H^2)r^2 \dot\Phi^2 }{4A^2F}
\end{align*}
where $E^{\m\n}=G^{\m\n}-T^{\m\n}$ are the Einstein equations \eqref{eoms}. Note that the left-hand sides of these equations are zero \textit{on shell}.} 
By comparing with \cite{Gregory:2018ghc}, one immediately sees that, in equivalent units $M_p^2 = 2$, our \eqref{eq2} corresponds to (47) in \cite{Gregory:2018ghc}. Furthermore, since \cite{Gregory:2018ghc} takes $A = 1$, equation \eqref{eq1} gives
\begin{equation}
    \frac{1-H^2}{F} = q(r) \Longrightarrow \partial_t H = -\frac{q(r)}{2 H} \partial_t F = \frac{r q(r) \dot{\Phi}^2}{4} = \frac{r (1-H^2)}{2 F} \frac{\dot{\Phi}^2}{2}
\end{equation}
which agrees with (48) in \cite{Gregory:2018ghc}. Lastly, our \eqref{eq3} corresponds to (49) in \cite{Gregory:2018ghc}.
\color{black}} to
\begin{eqnarray}
 2 H \partial_rA&=&\partial_t\left(\frac{1-H^2}{F}\right)\, ,\label{eq1}\\
\partial_t(r F)&=&-\frac{r^2H\dot \Phi^2}{2 A}\, ,\label{eq2}\\
\partial_r(r F)&=&1-\frac{V(\Phi)}{2}r^2-\frac{1-H^2}{A^2 F}\frac{r^2\dot\Phi^2}{4}\, ,\label{eq3}
\end{eqnarray}
where $\dot \Phi=\frac{d\Phi}{dt}$. One can verify that the three equations (\ref{eq1}-\ref{eq3}) imply all 10 Einstein equations, as well as the scalar equation of motion. 
\\
\\
Under the assumption $\partial_r A=0$, or equivalently $\partial_t\left(\frac{1-H^2}{F}\right)=0$, the above equations can be solved in full generality (see appendix \ref{notimeap}). Without this assumption, it is no longer possible to explicitly integrate the equations, at least as far as we are aware. Instead, we will discuss in section \ref{quasi-stationary expansion} how one can set up an approximation scheme of small (i.e. slow) $t$ dependence that allows to find a series solution whose coefficients can be iteratively solved for up to any required order.

\subsection{Time reparametrization invariance}\label{repsec}
By choosing the form of the metric and scalar field as (\ref{metricform}, \ref{scalarform}), we already removed almost all coordinate freedom. Still, the form (\ref{metricform}, \ref{scalarform}) remains invariant under reparametrizations of time\footnote{In the special case where $\Phi$ is constant a larger group of diffeomorphisms remains unfixed.}. I.e. a coordinate transformation of the form $t=t(\tilde t)$ under which the free functions transform as
\begin{equation}\label{general time transformation}
\tilde\Phi(\tilde t)=\Phi(t)\, ,\quad \tilde F(\tilde t,r)=F(t,r) \, ,\quad \tilde H(\tilde t,r)=H(t,r)\, ,\quad \tilde A(\tilde t)=\frac{dt}{d\tilde t} A(t)\, .
\end{equation}
For an infinitesimal such reparametrization $\delta t=-\Xi(t)$ this is equivalent to
\begin{equation}
\delta \Phi=\Xi \partial_t \Phi\, ,\quad \delta F=\Xi\partial_t F\, ,\quad \delta H=\Xi\partial_t H\, ,\quad \delta A=\Xi\partial_t A+A\partial_t \Xi\, .\label{deltau}
\end{equation}
One can explicitly verify that the Einstein equations in the form (\ref{eq1}-\ref{eq3}) do indeed remain invariant under these time reparametrizations. The transformations \eqref{deltau} will play a role in section \ref{timerep}, where they are used to simplify the expansion ansatz, upon which the remaining coordinate freedom will be fixed.

\subsection{Boundary conditions}\label{bcsec}
The equations (\ref{eq1}-\ref{eq3}) do not determine a unique solution. A priori, we have the unknown functions $A(r,t), H(r,t), F(r,t)$ and $\Phi(t)$ but since $H$ and $A$ only appear algebraically in (\ref{eq2}, \ref{eq3}), they can be eliminated in terms of $F,\Phi$ and their derivatives. This procedure then transforms \eqref{eq1} into a (non-linear) second-order PDE for $F$, once $\Phi$ is given. I.e. \ one is free to choose $\Phi(t)$ arbitrarily and only then $F$ can be solved for. Note that this is as expected, since -- as we discussed in the previous subsection -- a choice of $\Phi(t)$ can be interpreted as a choice of time coordinate and is hence a gauge freedom. Given $\Phi$, a solution to the PDE for $F$ will be uniquely determined only upon providing additional boundary conditions. Here we will postulate appropriate boundary conditions, following \cite{Gregory:2018ghc}, that impose regularity of the scalar field. Upon explicit solution of the problem using the quasi-stationary expansion in section \ref{quasi-stationary expansion}, we will see that the boundary conditions we introduce here do indeed provide unique and physically sensible solutions. 
\\
\\
Our interest is in solutions that generalize the static SdS solution by allowing time dependence, and as such we require some key features of SdS to remain present. Concretely, in addition to the assumption of spherical symmetry, we also demand there to be two horizons, the smallest of which we interpret as the black hole horizon while the largest one corresponds to the cosmological horizon. Let us recall, however, see e.g. \cite{Faraoni2015}, that for non-stationary geometries the notion of a horizon is no longer unique. The notions of event horizon and apparent horizon do now no longer coincide. As we show in appendix \ref{bcapp}, the apparent horizons of a metric of the form \eqref{metricform} are located -- as expected -- at the zeros of $F$. We'll assume\footnote{This is an assumption in this section, but will indeed follow from the Einstein equations when we will (approximately) solve them in section \ref{solsec}.} $F$ has two positive zeros, one at $r=\rb(t)$ and another at $r=\rc(t)$, and we will only be concerned with the solution in the region in between both horizons, i.e. $0\leq \rb(t)\leq r\leq \rc(t)$. 
\\
\\The physical boundary conditions we want to impose are then that the scalar field should be purely ingoing (from the perspective of the region between the horizons) on both apparent horizons. Or said differently, we want the horizons to be purely absorbing so that disturbances in the field can only leave the region between the horizons, not enter them, as time $t$ progresses. Although phrasing the boundary conditions like this makes them appear as a condition on the scalar field, they are in our setup a condition on the metric since we partially gaugefixed the coordinate dependence of the scalar field. Due to our gauge choice $\partial_r\Phi=0$, the requirement of the scalar field to be ingoing imposes that the metric should be such that $\pm F\partial_r$ is a future-oriented null vector that points into the respective horizon. We will explain this in more detail in appendix \ref{bcapp}, where we also show that this condition is equivalent to the boundary conditions
\begin{equation}
H(\rb(t),t)=1\quad\mbox{and}\quad H(\rc(t),t)=-1\,. \label{bc}
\end{equation}
For some intuition, we refer to the right diagram in Figure \ref{Penrose}, where $\partial_r$ is tangent to the purple lines of constant $t$. This happens because the equivalent of \eqref{bc}, see \eqref{boundary H}, holds for the SdS metric in $(t,r)$ coordinates.
\\ Although our derivation is somewhat different, the boundary conditions \eqref{bc} are essentially the same as those imposed in \cite{Gregory:2018ghc} (see their eqn (25)).

\section{The quasi-stationary expansion}\label{quasi-stationary expansion}
In this section, we introduce a small dimensionless parameter $\epsilon$, so that the Einstein equations can be solved order by order in an expansion of this small parameter. In this paper, we will do this explicitly for the first two orders in section \ref{solsec}. That the parameter $\epsilon$ is small amounts physically to the regime of slow time evolution. This is implemented by assuming the time coordinate $t$ only appears in the fields through the combination $\tau=\epsilon t$, which implies all time derivatives are suppressed by a factor $\epsilon$. As such, when time derivatives-- and thus velocities-- are small, one expects to be in the nonrelativistic regime. Indeed this is the case; as we show in detail in section \ref{1csec}, upon the identification $\epsilon\sim\frac{1}{c}$, the expansion in $\epsilon$ becomes equivalent to the nonrelativistic $1/c$ expansion. We refer to \cite{Hartong2022} for an introduction. 
\\ As explained below, we will simultaneously assume the scalar field only appears in the potential through $\varphi(\tau)=\sqrt{\epsilon}\Phi(t)$. This guarantees the potential is shallow enough to accommodate slow time evolution, no matter how small $\epsilon$. The expansion introduced below implies the usual slow-roll approximation, through $(V'(\Phi))^2\sim V''(\Phi)\sim \epsilon^2$, with $\epsilon$ providing a controllable parameter that allows to compute corrections order by order.
\\At leading order, the Einstein equations will become those of a stationary metric ansatz, so that the solution is a quasi-stationary solution. Here, the integration constants -- that are indeed constants for a stationary solution -- are now functions of the slow time parameter $\tau$. For this reason, we will refer to the expansion in $\epsilon$ as the quasi-stationary expansion.

\subsection{Setup}\label{setup}
We construct a slow time expansion with non-trivial dynamics in two steps. First, it is assumed that the time dependence only appears through the slow time variable $\tau =\epsilon\, t$. This is so for all the functions we consider, but in particular for the scalar field, so that $\Phi(t)=\phi(\epsilon t)$. This assumption amounts to a replacement of all $\partial_t$ by $\epsilon\partial_\tau$ and $\Phi$ by $\phi$. The scalar equation then takes the schematic form
\begin{equation}
\epsilon^2\mathbb{A}\partial_\tau(\mathbb{B}\partial_\tau\phi)+\e\, \mathbb{C}\partial_\tau \phi= -\partial_\phi V(\phi)\, ,\label{scaleq}
\end{equation}
where $\mathbb{A}$, $\mathbb{B}$ and $\mathbb{C}$ depend on the metric coefficients. One sees that for small $\epsilon$ both terms with a time derivative get suppressed. But to allow non-trivial dynamics\footnote{In the one-scalar model we consider, the equation $V'=0$ has only a discrete set of solutions. One could consider a multi-scalar model that has a continuous set of solutions to $V'=0$, in which case the usual slow time dynamics would take place on that set. But such dynamics would, by definition, have $\partial_\tau V=0$ and this, in turn, implies a time-independent cosmological constant. As we are interested in a dynamical cosmological constant we need to consider non-zero $V'$, which then will source $\phi$.} we need the $\partial_\phi V$ term to at least source the $\partial_\tau\phi$ term. In the second step, this is achieved by redefining\footnote{One easily checks that in redefining $\phi=\epsilon^a\varphi$ only $a=-\frac{1}{2}$ provides the desired scalings.} $\varphi=\sqrt{\epsilon}\phi$ and assuming the potential is only a function of the rescaled scalar field $V(\phi)=v(\sqrt{\epsilon}\phi)$ so that the scalar equation becomes
\begin{equation}
\epsilon\mathbb{A}\partial_\tau(\mathbb{B}\partial_\tau\varphi)+\mathbb{C}\partial_\tau \varphi=-\partial_\varphi v(\varphi)\, .\label{scaleqschem}
\end{equation}
In the limit $\epsilon\rightarrow 0$,  we lose the $\partial_\tau^2\varphi$ term but keep a non-trivial equation for $\partial_\tau\varphi$. At leading order, the dynamics are thus friction dominated and described by a gradient flow, with corrections due to the term second order in time derivatives only appearing at higher order. To gain some intuition about such an expansion we have added a short discussion of an identical expansion applied to the mechanics of a damped oscillator in appendix \ref{toyap}.  
\\
\\
In summary, then, the above argument suggests the introduction of a small parameter $\epsilon$ into the metric and scalar field (\ref{metricform}, \ref{scalarform}) as
\begin{align}
F(t,r)=f(\epsilon t,r)\,,\qquad &H(t,r)=\eta(\epsilon t,r)\,,\qquad A(t,r)=a(\epsilon t,r)\, ,\nonumber\\ \sqrt{\epsilon}\,\Phi(t)=\varphi(\epsilon t)\,,\qquad& V(\Phi)=v(\sqrt{\epsilon}\Phi)\,,\qquad \tau=\epsilon t\,. \label{smallparam}
\end{align}
In terms of these new variables the Einstein equations (\ref{eq1},\ref{eq3}) are then equivalent\footnote{Assuming $\frac{d}{d\tau}\varphi\neq 0$.} to
\begin{eqnarray}
\zeta \partial_r b&=&\epsilon\frac{d}{d\tau}\frac{b \zeta^2-1}{f}\, ,\label{feq1}\\
\partial_r\left(r(f-1+\frac{v}{6}r^2)\right)&=&\epsilon \frac{b \zeta^2-1}{bf}\frac{r^2\dot\varphi^2}{4}\, .\label{feq2}
\end{eqnarray}
where $\dot\varphi=\frac{d}{d\tau}\varphi$ and
\begin{equation}
b=a^2\,,\qquad \zeta=-\frac{\eta}{a}\, .	\label{bzetadef}
\end{equation}
Furthermore $\zeta$ is determined, via \eqref{eq2}, algebraically in terms of $f$ and $\dot\varphi$ as
\begin{equation}
\zeta=\frac{2\, \partial_\tau f}{r \dot \varphi^2}\, .\label{zetadef}
\end{equation}
{We want to emphasize here that, in what follows, only the case $\dot{\varphi} \neq 0$ is considered as we are interested only in dynamics solutions. Of course, $\dot\varphi$ could dynamically evolve towards zero, at which point our analysis would break down. We comment shortly on this at the end of subsection \ref{behaviour of the solutions}.}

Recall { now} that the scalar equation \eqref{scaleqschem} is automatically implied by (\ref{feq1}, \ref{feq2}, \ref{zetadef}).
After the introduction of $\epsilon$ and our final rewriting, we are left with the unknown functions $f(\tau,r)$, $b(\tau,r)$, $\varphi(\tau)$ which are constrained only by the equations (\ref{feq1}, \ref{feq2}) and the boundary conditions \eqref{bc} which become
\begin{equation}
\left.\left(2 \sqrt{b}\, \partial_\tau f\right)\right|_{r=\rb}=-\rb \dot \varphi^2\, ,\quad\quad \left.\left(2 \sqrt{b}\, \partial_\tau f\right)\right|_{r=\rc}=\rc \dot \varphi^2 \, ,\label{bc2}
\end{equation}
where the roots $0\leq \rb\leq \rc$ are zeros of $f$.
\\
\\
As is clear by observation, and as we will see more explicitly below, the equations (\ref{feq1}, \ref{feq2}) behave very well in the $\epsilon\rightarrow 0$ limit and $\epsilon$ appears in a simple linear fashion. This suggests assuming that the fields are analytic in $\epsilon$ and then solving the equations perturbatively in this parameter. This suggests the ansatz
\begin{equation}
f(\tau,r)=\sum_{n=0}^\infty f_n(\tau,r) \epsilon^n\, ,\quad b(\tau,r)=\sum_{n=0}^\infty b_n(\tau,r) \epsilon^n\, ,\quad \varphi(\tau)=\sum_{n=0}^\infty \varphi_n(\tau) \epsilon^n\, . \label{slowtimexp}
\end{equation}
It then follows that all derived quantities such as $a$, $\zeta$, $\eta$, $\rb$, $\rc$, as well as the metric coefficients $g_{\m\n}$ themselves, will be analytic in $\epsilon$, and their expansion is straightforwardly determined in terms of that of the `fundamental' variables $f, b$ and $\varphi$. Let us point out that a priori one could assume $v$ to depend on $\epsilon$ as well, i.e. $v=\sum_n v_n\epsilon^n$, but for simplicity, we'll assume this not to be the case, i.e. $v=v_0$. 
\\
\\
Lastly, as we will discuss in subsection \ref{timerep}, using an appropriate time reparametrization, one can choose the time parameter $\tau$ such that $b_0 = 1$ and $\varphi_n = 0 \ \forall n \geq 1$ without loss of generality. The only freedom left in the $\tau$ parameter is then a constant time translation $\tau \to \tau + \tau_0$, which manifests itself through the equations (\ref{feq1}, \ref{feq2}) being autonomous in $\tau$. As a result, due to the presence of symmetry in the system, one expects to find a conserved quantity in the dynamics, something which we will see explicitly in the next section.
\\
In conclusion, finding a solution of the Einstein equations for (\ref{metricform}, \ref{scalarform}) then amounts to solving (\ref{feq1}, \ref{feq2}) order by order for
\begin{equation*}
\varphi=\varphi_0\, , \qquad f_n\,,\ n=0,1,\ldots\, , \qquad b_n\,,\ n=1,2,\ldots\, .
\end{equation*}
We will do this for $n=0$ and $n=1$ in sections \ref{LO} and \ref{NLO} respectively. 
\subsection{Fixing time reparametrizations}\label{timerep}
As discussed in section \ref{repsec}, the formalism is invariant under time reparameterizations, which after the introduction of $\tau=\epsilon t$, equivalently take the form $\tau=\tau(\tilde \tau)$. A key observation is now that such a redefinition of $\tau$ can non-trivially depend on $\epsilon$. If we require that our functions remain analytic under a time reparametrization, then the function $\tau(\tilde\tau)$ has to be itself analytic in $\epsilon$. Doing such an $\epsilon$ dependent time reparametrization will reshuffle the coefficients in the series expansion of $f, b$ and $\varphi$. This is most easily tracked explicitly in infinitesimal form. Writing\footnote{I.e. $\epsilon\Xi(t)= -\epsilon \delta t = - \delta \tau = \xi(\epsilon t)$.} $\delta \tau=-\xi(\tau)$ and assuming\footnote{{For a more detailed discussion of the interplay between expansions of GR and diffeomorphism invariance, see e.g. \cite{Bleeken2017}.}} $\xi(\tau)=\sum_n \xi_n(\tau)\epsilon^n$ one finds that \eqref{deltau} becomes
\begin{equation}
\delta f_n=\sum_{k=0}^n \xi_k\partial_\tau f_{n-k}\,,\quad \delta b_n=\sum_{k=0}^n( \xi_k\partial_\tau b_{n-k} +2 b_{n-k}\partial_\tau\xi_k)\,,\quad \delta \varphi_n=\sum_{k=0}^n \xi_k\dot \varphi_{n-k}\label{deltatau}
\end{equation}
The upshot is that, since $f_n$ and $b_n$ are functions of both $\tau$ and $r$, they can't be simplified using time reparametrization since the $\xi_k$ only depend on $\tau$. However, because the $\varphi_n$ are functions only of $\tau$, they can be changed arbitrarily by a time reparametrization. Indeed, one could even choose the scalar field as being the time coordinate, which would be drastic and not very intuitive. But in the slow time expansion, it is a lot more natural to choose time such that $\varphi_n =0$ for all $n\geq 1$.
\\As one sees from \eqref{deltatau}, this fixes all $\xi_k$ for $k\geq 1$, but not $\xi_0$, i.e. there remains an $\epsilon$ independent time reparameterization. Rather than use this remaining freedom to put $\varphi_0$ to some preferred (non-constant) function, we decide to fix the $\epsilon$ independent reparametrizations in some other way. 
It turns out there is an excellent option since one sees directly from \eqref{feq1} that, although $b_n$ will be a function of $\tau$ and $r$ when $n\geq 0$, the leading coefficient, $b_0$, depends on time only. As such, we use the leading order time reparametrization invariance $\xi_0$ to put $b_0$, to a constant, say 1. This removes the freedom to shift by $\xi_0$ and then fix all time-reparametrizations (up to constant time shifts).
\\
\\
In summary, as we already anticipated at the end of section \ref{setup}, we can always choose our time $\tau$ such that
\begin{equation}
b_0=1\,,\qquad \varphi=\varphi_0\quad\Leftrightarrow\quad \varphi_n=0\quad \forall n\geq 1\label{b0is1}
\end{equation}
and upon this choice, the only symmetry left are constant shifts of time $\tilde\tau=\tau+\tau_0$.

\subsection{The quasi-stationary setup as a nonrelativistic expansion}\label{1csec}
By the construction in section \ref{setup}, for small $\epsilon$, all the time derivatives are suppressed such that the quasi-stationary setup can alternatively be viewed as a small velocity and hence nonrelativistic $1/c$ expansion, with $c$ the speed of light. The $1/c$ setup, see \cite{Bleeken2017, Hansen2020, Hartong2022} for a review, is a generalization of the more commonly used Post-Newtonian (PN) expansion. Indeed, where the PN setup additionally assumes weak gravitational fields (both small $1/c$ and $G_\mathrm{N}$ limit), this is not the case in the $1/c$ expansion, which is a purely non-relativistic expansion where, at leading order, the metric can have arbitrarily high curvature. 
\\ In this subsection, we explicitly point out the equivalence of the quasistationary $\epsilon$ and nonrelativistic $1/c$ expansion. It was already pointed out in \cite{Bleeken2019, Ergen2020}, and further made precise in \cite{Elbistan2022}, that the $1/c$ setup can be considered as an expansion in time derivatives around a quasi-stationary background. As such, the interchangeability of the $\epsilon$ and $1/c$ expansions is no surprise. Still, spelling out this equivalence has its merit, since to our knowledge, in the literature on the $1/c$ setup, all examples so far have been expansions of exact solutions of the Einstein equations. The solution up to the next-to-leading order we will discuss in the next section is one for which no all-order closed-form solution is known. As such, the solution of this paper is also the first example of an application of the $1/c$ expansion beyond the exact solutions of GR.
\\
\\
Introducing the speed of light into the metric \eqref{metricform} gives the form
\begin{equation*}
ds^2=-c^2 A^2  F dt^2+2 c A H dt dr+\frac{1-H^2}{F}dr^2+r^2 d\Omega^2\, ,
\end{equation*}
where it is now explicit that $t$ has dimensions of time. Rewriting the above in terms of the time $\tau=\epsilon t$ and using the redefinitions \eqref{smallparam} one gets
\begin{equation}
ds^2=-\left(\frac{c}{\epsilon}\right)^2 a^2  f d\tau^2+2 \frac{c}{\epsilon} a \eta d\tau dr+\frac{1-\eta^2}{f}dr^2+r^2 d\Omega^2\, .\label{sphsymKS}
\end{equation}
The key observation is that the speed of light $c$ now appears in the metric only through the combination $\mathbf{c}=\frac{c}{\epsilon}$. Assuming $\epsilon$ to be small is equivalent to assuming $\mathbf{c}$ to be large, which is exactly the assumption of the $1/c$ expansion, where the speed of light is treated as a free parameter\footnote{The $1/c$ expansion is formal in that $c$ is of course a dimensionful quantity. In particular examples and applications, the small parameter in the expansion is $\frac{v}{c}$ with $v\ll c$ some typical velocity. If one wants to introduce a tuneable parameter in the expansion that is arbitrarily small, one can rescale $v\rightarrow \epsilon v$ so that the small dimensionless parameter becomes $\frac{v}{\mathbf{c}}=\epsilon\frac{v}{c}$. But since the physical interpretation and origin of $v$ can differ from application to application it is convenient to simply work with $c$ itself as a formal expansion parameter.} that can be made arbitrarily large. In terms of this parametric speed of light $\mathbf{c}$, the metric ansatz of \cite{Elbistan2022} takes the form
\begin{equation*}
ds^2=-e^\psi (\mathbf{c}\,dt_{\mathrm{KS}}+C_i dx^i)^2+e^{-\psi}\gamma_{ij}dx^i dx^j\, .
\end{equation*}
One sees that upon the identification $t_{\mathrm{KS}}=\tau$, the metric \eqref{sphsymKS} is a special case of the metric above where
\begin{equation*}
e^\psi=a^2 f\, ,\qquad C_idx^i=-\frac{\eta}{a f}dr\, ,\qquad \gamma_{ij}dx^idx^j=a^2(dr^2+f r^2 d\Omega^2)\, .
\end{equation*}
In the $1/c$ setup of \cite{Elbistan2022}, one makes the expansion ansatz
\begin{equation*}
\psi=\sum_{n=0}^\infty \os{\psi}{n}\,\mathbf{c}^{-n}\, ,\qquad C_i =\sum_{n=0}^\infty \os{C}{n}_i\,\mathbf{c}^{-n}\, ,\qquad \gamma_{ij}=\sum_{n=0}^\infty \os{\g}{n}_{ij}\,\mathbf{c}^{-n}\, .
\end{equation*}
which, upon setting $c=1$ so that $\mathbf{c}^{-1}=\epsilon$, can be identified with the quasi-stationary expansion in $\epsilon$ as introduced in section \ref{setup}. The only difference is then in the choice of variables, which, up to NLO, are related as
\begin{align}
\os{\psi}{0}&=\log f_0\, , \qquad\qquad\qquad\qquad\quad\,\os{\psi}{1}=b_1+\frac{f_1}{f_0}\, ,\nonumber\\
\os{C}{0}_i dx^i&=\frac{\zeta_0}{f_0}dr\, ,\qquad\qquad\qquad\quad\ \ \os{C}{1}_i dx^i= \frac{f_0\zeta_1-f_1\zeta_0}{f_0^2}dr\, ,\label{id}\\
\os{\gamma}{0}_{ij}dx^idx^j&=dr^2+f_0 r^2 d\Omega^2\, ,\qquad \os{\gamma}{1}_{ij}dx^idx^j=b_1dr^2+(b_1f_0+f_1) r^2 d\Omega^2 \, .\nonumber
\end{align}
where we used \eqref{bzetadef} and the coordinate choice $b_0=1$. This precise identification of variables can be continued to arbitrary order $\mathbf{c}^{-n}=\epsilon^{n}$, but \eqref{id} is sufficient to translate the results of section \ref{solsec} to solutions of the NLO expanded Einstein equations as formulated in \cite{Elbistan2022}. This mapping in turn can be translated into the more geometric Newton-Cartan formulation used in \cite{Bleeken2017, Hansen2020}.
\section{Approximate solution for small \texorpdfstring{$\epsilon$}{ε}}\label{solsec}
This section explicitly determines the solution up to the first subleading order in the quasi-stationary setup. The expansion starts at the leading order (LO) $\calo(\epsilon^0) = \calo(1)$, while the next-to-leading order (NLO) corresponds to $\calo(\epsilon^1)$. This means that the solution obtained by combining the LO and NLO results below will solve all equations up to an error of $\calo(\epsilon^2)$. At this order of approximation, the metric takes the form\footnote{ Remember that we made the choice $b_0=1$, see \eqref{b0is1}. That this choice is always possible is elaborated on in section \ref{LO}.}
\begin{align}\label{full lo + nlo metric}
ds^2=&-\left(f_0+\epsilon(b_1 f_0+f_1)+\calo(\epsilon^2)\right)dt^2-2\left(\zeta_0+\epsilon(b_1\zeta_0+\zeta_1)+\calo(\epsilon^2)\right)dt dr\nonumber\\
&+\left(\frac{1-\zeta_0^2}{f_0}-\epsilon \frac{(1-\zeta_0^2)f_1+(b_1\zeta_0+2\zeta_1)\zeta_0 f_0}{f_0^2}+\calo(\epsilon^2)\right)dr^2 +r^2 d\Omega^2
\end{align}
The expression for $f_0$ can be found in \eqref{tdwarp}, while those for $f_1$ and $b_1$ are given in \eqref{fb1}. Note that these formulas contain various $\tau$ dependent functions -- including the scalar field $\varphi(\tau)$ -- and the ODEs determining these functions, as well as the method to solve those, are discussed in detail below. Furthermore, let us recall from \eqref{zetadef}  that $\zeta_n= \frac{2 \  \partial_\tau f_n}{r\dot\varphi^2}$; an explicit expression for $\zeta_0$ is given in \eqref{zeta0}. Finally we remark that $dt=\frac{d\tau}{\epsilon}=\mathbf{c}dt_\mathrm{KS}$ as was discussed in section \ref{1csec}.

\subsection{The solution at leading order}\label{LO}
The Einstein equations at leading order are obtained simply by putting $\epsilon=0$ in (\ref{feq1}, \ref{feq2}), which gives  
\begin{eqnarray}
\zeta_0 \partial_r b_0&=&0\, , \label{b0eq}\\
\partial_r\left(r(f_0-1+\frac{v}{6}r^2)\right)&=&0\, . \label{f0eq}
\end{eqnarray}
\subsubsection{Solving the equations}\label{solving equations lo}
The first equation implies $b_0$ is $r$ independent and hence a function of $\tau$ only. As we anticipated in section \ref{timerep}, this implies we can -- without loss of generality -- use the time-reparametrization invariance to put
\begin{equation*}
b_0=1\,.
\end{equation*}
The remaining equation \eqref{f0eq} is directly integrated to 
\begin{equation}
f_0(r,\tau)=1-\frac{2m(\tau)}{r}-\frac{r^2}{3\ell^2(\tau)}\,,\label{tdwarp}
\end{equation}
where
\begin{equation}
\ell^2(\tau)=\frac{2}{v(\varphi(\tau))}\label{fixfi}\,.
\end{equation}
We recognize in \eqref{tdwarp} the warpfactor of the SdS metric, see \eqref{fdef}, but where the mass $m(\tau)$ and cosmic length scale $\ell(\tau)$ are now two\footnote{As we'll see below, it is convenient to choose $\ell(\tau)$ as an independent function to be determined, and $\varphi(\tau)$ as a dependent variable, defined algebraically in terms of $\ell(\tau)$ via \eqref{fixfi} through a choice of scalar potential $v(\varphi)$. } arbitrary functions of time that remain to be determined.
\\
\\
Since (\ref{b0eq}, \ref{f0eq}) imply the full Einstein and scalar equations of motion at leading order, we see here explicitly that those do not fix the time dependence of the solution fully. As we will now discuss, the physical time dependence -- or dynamics -- of the solution is ultimately set by imposing the boundary conditions \eqref{bc2}, which translate into two coupled first-order ODEs for $m(\tau)$ and $\ell(\tau)$.
\\
At leading order, or equivalently in the limit $\epsilon\rightarrow 0$, the boundary conditions \eqref{bc2} become
\begin{equation}
\left.\left(2 \partial_\tau f_0\right)\right|_{r=\rbz}=-\rbz \dot \varphi^2\, ,\quad\quad \left.\left(2\, \partial_\tau f_0\right)\right|_{r=\rcz}=\rcz \dot \varphi^2\, . \label{bc0}
\end{equation}
Here $\rbz, \rcz$ are the order $\calo(\epsilon^0)$ terms in the $\epsilon$ expansion of $\rb, \rc$, or equivalently the positive roots of $f_0$. Since $f_0$ has the same form as in the section
\ref{SdSsection}, it follows we have expressions identical to (\ref{adef}-\ref{rcdef}), but with $m$, or $\alpha$, and $\ell$ functions of $\tau$. If we define $\alpha(\tau)$ through
\begin{equation*}
m(\tau)=\frac{\ell(\tau)\cos\left( 3\alpha(\tau)\right)}{3}\, .\label{ataudef}
\end{equation*}
Then
\begin{eqnarray*}
\rbz(\tau)&=&2\ell(\tau)\cos\left(\a(\tau)+\frac{\pi}{3}\right)\, ,\\ \rcz(\tau)&=&2\ell(\tau)\cos\left(\a(\tau)-\frac{\pi}{3}\right)\,.\label{rczdef}
\end{eqnarray*}
From now on, the variables $m, \ell, \alpha, \varphi, \rbz$ and $\rcz$ should be interpreted as functions of $\tau$ and we will no longer write this dependence explicitly. The boundary conditions \eqref{bc0} are then equivalent to the pair of coupled ODEs
\begin{equation}
\dot{\alpha}=\frac{\ell \dot{\varphi}^2}{2\sqrt{3}}\frac{\sin\a}{\tan 3\a}\, ,\qquad \dot \ell=\frac{\ell^2 \dot{\varphi}^2}{2\sqrt{3}}\frac{1-\frac{1}{2}\cos 2\a}{\sin\a}\, .\label{ODEs}
\end{equation}
These can be solved up to quadrature as follows. First, one remarks that there is a conserved quantity
\begin{equation}
M=(12)^{-\frac{3}{4}}\frac{\cos 3\alpha}{\sin\a\sqrt{\sin 2\a}}\ell\,,\label{Mdef}
\end{equation}
i.e. \eqref{ODEs} imply that $\frac{d M}{d\tau}=0$. This is to be expected from Noethers theorem and the remaining constant time shift symmetry $\Tilde{\tau} = \tau + \tau_0$. The above can be rewritten as an expression for $\ell$ in terms of $M$ and $\alpha$, such that
\begin{equation}
\hspace{-1 cm} \ell(\tau)=(12)^{\frac{3}{4}}\frac{\sin\a(\tau)\sqrt{\sin\left(2\alpha(\tau)\right)}}{\cos \left(3\a(\tau)\right)}M\quad \mbox{and}\quad  m(\tau)=\frac{2^{\frac{3}{2}}}{3^{\frac{1}{4}}}M\sin\alpha(\tau)\sqrt{\sin \left(2\a(\tau)\right)}\,.\label{ltau}
\end{equation}
This matches and motivates the $(M,\alpha)$ parametrization we already introduced in section \ref{SdSsection}, since $M$ is constant and time dependence enters exclusively through $\alpha$.
The numeric prefactor in \eqref{Mdef} is chosen such that $\left.m\right|_{\alpha=\frac{\pi}{6}}=M$. There remains then a single ODE determining $\alpha(\tau)$: 
\begin{equation}\label{ode alpha}
\dot\alpha=\frac{\sin 3\alpha}{3^{\frac{1}{4}}M\left(\frac{d\varphi}{d\alpha}\right)^{2}\sqrt{2\sin 2\alpha}\,\sin^2\alpha} \, ,
\end{equation}
where $\frac{d\varphi}{d\alpha}$ is a function of $\alpha$ determined through the inverse of the scalar potential via
\begin{equation}
\varphi(\alpha)=v^{-1}\left(\frac{2}{\ell^2(\alpha)}\right)\,.\label{phialpha}
\end{equation}
Here $\ell(\alpha)$ is to be interpreted as the leftmost expression in \eqref{ltau} with $\alpha(\tau)$ replaced by $\alpha$. Once a scalar potential $v$ is specified, \eqref{ode alpha} can then be solved by integration for $\tau(\alpha)$. Remark that the full solution then has two integration constants, $M$ and $\alpha(0)$, and thus -- just like the static SdS metrics -- forms part of a two-parameter family of solutions.
\\
{Finally, through \eqref{zetadef} and the discussion of the ODEs above, we find the last metric component $\zeta_0$:
\begin{align}
\zeta_0(r,\tau)&=-\eta_0(r,\tau)=-\frac{\beta(\tau)}{r^2}+\gamma(\tau)\,r\label{zeta0}\\
\beta&=\left(4M\sin\a\right)^2 \qquad \gamma =\frac{(1+2\sin^2\a)\cos 3\a}{(108)^{\frac{3}{4}}M \sin^2\a\,\sqrt{\sin 2\a}} \label{beta gamma}
\end{align}
where $\alpha(\tau)$ is determined via \eqref{ode alpha}.}
\subsubsection{Dynamic behaviour of the solutions}\label{behaviour of the solutions}
It turns out that the generic features of the dynamics are independent of the choice of potential, as long as it is everywhere non-negative, i.e. $v(\varphi)\geq 0$. Let us now assume\footnote{Recall from section \ref{SdSsection} that one only has a physical solution when $\ell> 0$ and $0\leq \alpha\leq \frac{\pi}{6}$, which implies $M>0$. But remark that in the $(M,\alpha)$ parametrization, keeping $M$ fixed and finite, the limit $\alpha=0$ corresponds to an infinite cosmological constant, which is unphysical. So we exclude it from our starting assumption. On the other hand, the case $\alpha=\frac{\pi}{6}$, which corresponds to the asymptotically flat Schwarzschild black holes, is of course very physical. However, it is not an interesting starting point, since it corresponds to $v=0$, which is a minimum for any positive potential, and it would lead to a static solution.} that the dynamics start in the region $ M>0, 0< \alpha< \frac{\pi}{6}$ at $\tau=0$. One then sees that always $\dot{\alpha}\geq 0$ via \eqref{ode alpha}, so $\alpha$ can only increase under time evolution. This remains true as long as $0<\alpha<\frac{\pi}{6}$ and we will show later that, for a large class of scalar field potentials, the dynamics automatically ends if $\alpha$ reaches the end point of validity $\frac{\pi}{6}$. 
\\ Furthermore, it is straightforward to compute that
\begin{eqnarray}
\frac{d\ell}{d\alpha}&=&2M \,3^{3/4}\frac{(1+2\sin^2\alpha)\sin 3\a}{\cos^2 3\a\sqrt{\tan\a}}\geq 0 \nonumber\, , \\
\frac{dm}{d\alpha}&=&\frac{2^{\frac{3}{2}}M\sin 3\a}{3^{\frac{1}{4}}\sqrt{\sin 2\a}}\geq 0 \nonumber\, , \\
\frac{dv}{d\alpha}&=&-\frac{(1+2\sin^2\alpha)\sin 6\alpha}{24\sqrt{3}M \sin^4\alpha \sin 2\a }\leq 0\, .\label{vp}
\end{eqnarray}
Since $\alpha$ is monotonically increasing under time evolution we can then conclude the following three generic features of the dynamics:
\begin{itemize}
	\item $\ell$ increases, $\Lambda$ decreases
	\item $v$ decreases
	\item $m$ increases
\end{itemize} 
This is indeed what one would expect on physical grounds, in that it matches the intuition of the scalar field rolling down the potential and thereby decreasing the value of the cosmological constant. Furthermore, in this process, the black hole absorbs the scalar field and grows, {i.e. its mass $m$ increases}.
\\
\\
{Finally let us comment on the endpoint of the evolution. As remarked above, the potential is assumed to be positive and its value continuously decreases. This appears to imply that the dynamics will have to come to a halt, i.e. $\dot\varphi=0$, when the minimum of the potential is reached. Although this is indeed the case for a large class of potentials, there are some exceptions. The loophole in the previous argument is that it is based on the assumption $\alpha<\frac{\pi}{6}$ we made earlier. As explained, this condition can, without loss of generality, be chosen for the initial state, but $\alpha=\frac{\pi}{6}$ could be reached dynamically. That this leads to an exception is most easily seen from the scalar equation of motion. Working out \eqref{scaleqschem} explicitly at leading order, $\calo(\epsilon^0)$, gives:
	\begin{equation}
		3 \gamma(\alpha(\tau)) \cdot \dot{\varphi} = - \partial_\varphi v(\varphi)\label{scalfric}
	\end{equation}
where $\gamma$ is given in \eqref{beta gamma}. 
It follows from this equation that $\dot\varphi$ has to vanish at the minimum of the potential, i.e. $\partial_\varphi v = 0$, except when $\gamma=0$.
Looking at \eqref{beta gamma} one sees that $\gamma=0 \Leftrightarrow \alpha=\frac{\pi}{6}$. We thus recover the conclusion 
that the dynamics will evolve towards the minimum of the potential and stop there, except for the cases where $\alpha$ reaches $\frac{\pi}{6}$. What happens at $\alpha = \frac{\pi}{6}$, i.e. whether the particle still stops at or overshoots the minimum, depends on the potential and needs to be studied case by case. 
\\Let us briefly discuss the physics behind this behaviour. Equation \eqref{scalfric} represents an overdamped particle rolling in a potential $v(\varphi)$ with time-dependent friction $\gamma(\alpha(\tau))$. Unless this damping vanishes, the particle will come to rest at the minimum of the potential. In our gravitational setting, the friction is the well-known Hubble friction related to cosmological expansion and indeed, 
the case $\alpha=\frac{\pi}{6}$ corresponds to the $\Lambda=0$ Schwarzschild black holes, i.e. the only black holes in the SdS class for which the expansion, and hence Hubble friction, vanishes. 
\\
\\
Apart from the behaviour of $\varphi$ when it reaches the minimum of the potential, i.e. stopping or shooting through, there is also the issue of at what time $\tau$ the minimum is reached. Studying a few example potentials reveals that this can happen asymptotically, i.e. as $\tau\rightarrow \infty$, but sometimes also at a finite time. 
\\In case the minimum is reached at finite time, say $\tau=\tau_\star$, one has to worry about what happens to the space-time at that point. 
Based on our previous discussion, apart from the special case $\alpha=\frac{\pi}{6}$, the point $\dot\varphi=0$ will be reached after a finite time $\tau_\star$.
This is of relevance since it means that, at $\tau=\tau_\star$, the solution evolves outside the following assumptions: 
$\dot\varphi\neq0$ to make the gauge choice \eqref{scalarform} and $\zeta$ being well-defined in \eqref{zetadef} when solving the equations of motion. Optimistically, 
at $\tau=\tau_\star$, one can simply glue the dynamic solution obtained for $0\leq\tau<\tau_\star$ onto the corresponding stationary SdS solution for $\tau_\star\leq \tau$. Investigating this in more detail would require an analysis based on the Israel matching conditions \cite{Israel:1966rt} and falls outside the scope of this paper. 
\\It is of course not \textit{a priory} excluded that, in certain cases, a singularity develops rather than a smooth continuation into a stationary solution that one would hope for. Let us stress however that for a large class of potentials, e.g. those of the form $V(\varphi)=\varphi^{-\nu}$, $\nu>1$, this issue of breakdown/continuation at finite time does not arise. Indeed, in these cases $\dot\varphi>0$ for all $\tau\in[0,\infty)$ and the solution remains inside the domain of validity of our analysis at all times. Studying in more detail the end phase of the dynamics, its relation to the choice of scalar potential and the corresponding regime of validity of the methods of this paper are interesting avenues for further research.
}

\subsubsection{Comparison with \cite{Gregory:2018ghc} \color{black}}\label{comparison with 1}
In this subsection, we compare our solution at leading order to the static SdS space-time and to the leading order\footnote{In our setup, the leading order corresponds to focussing on the lowest power of $\epsilon$ in the expansion \color{black}, i.e. $\calo(\epsilon^0)$, which is equivalent to dropping all time derivatives in the equations.  This translates to the setup of \cite{Gregory:2018ghc} by dropping the terms $\delta f$ and $\delta \eta$, since these are of order $\calo(\epsilon)$\color{black}. However, there \textit{does} remains a non-trivial time-dependence in the metric components. 
 \color{black} } part of the metric of \cite{Gregory:2018ghc}. 

All three metrics have identical radial dependence and share the form
\begin{equation}
ds^2=-f dt^2+2\eta dt dr+\frac{1-\eta^2}{f}dr^2+r^2 d\Omega_2^2\, ,
\end{equation}
where
\begin{equation}
	f(r,t)=1-\frac{2 m(t) }{r}-\frac{\Lambda(t)}{6}r^2\qquad \eta(r,t)=\frac{\beta(t)}{r^2}-\gamma(t)r
\end{equation}
The difference between the metrics is in their time dependence, as we now spell out.

\paragraph{Schwarzschild-de Sitter}
This metric is static (albeit in `stationary form' here), which corresponds to
\begin{equation}
	f(r,t)=f_{\mathrm{SdS}}(r)=1-\frac{2 m }{r}-\frac{\Lambda}{6}r^2\qquad \eta(r,t)=\eta_{\mathrm{SdS}}(r)=\frac{\beta}{r^2}-\gamma r
\end{equation}
In the above $m, \Lambda$ and $\beta, \gamma$ are real numbers, i.e. constants. Recall that $\beta$ and $\gamma$ are themselves precisely determined in terms of $m$ and $\Lambda$, see \eqref{Hform}.

\paragraph{Leading order metric of \cite{Gregory:2018ghc}}
Reading from equations (38, 39) in \cite{Gregory:2018ghc}, their leading order metric corresponds to
\begin{equation}
	f(r,t)=f_{\mathrm{QS}}(r,t)=1-\frac{2 m(  t) }{r}-\frac{\Lambda(  t)}{6}r^2\qquad \eta(r,t)=\eta_{\mathrm{SdS}}(r)=\frac{\beta}{r^2}-\gamma r \label{fetheirs}
\end{equation}
We see that this corresponds to a `partially' quasi-stationary version of SdS, in that time dependence is introduced in $f$ through the integration constants $m, \Lambda$ of the radial problem only, but 
not in $\eta$. This is peculiar since $\beta$ and $\gamma$ are not independent integration constants, but rather themselves functions of $m$ and $\Lambda$. In other words, the leading order metric of \cite{Gregory:2018ghc} corresponds to a quasi-stationary replacement of $(m,\Lambda)\rightarrow (m(  t),\Lambda(  t))$ in the $dt^2$ component of the metric, but not at all in the $dt dr$ component and only partially in the $dr^2$ component.

\paragraph{Leading order metric in this paper}
Through our setup with a precise expansion in the small parameter $\epsilon$, we found a leading order metric which corresponds to
\begin{equation}
	f(r,t)=f_0(r,t)=1-\frac{2 m(   t) }{r}-\frac{\Lambda(   t)}{6}r^2\qquad \eta(r,t)=-\zeta_{0}(r,t)=\frac{\beta(   t)}{r^2}-\gamma(   t) r \label{feours}
\end{equation}
Furthermore the time dependence of $\beta$ and $\gamma$ is such that one takes the time independent SdS result $\beta(m,\Lambda)$ and $\gamma(m,\Lambda)$ and replaces the integration constants $m,\Lambda$ with their time-dependent versions: $\beta(   t)=\beta(m(  t),\Lambda(  t))$, $\gamma(  t)=\gamma(m(  t),\Lambda(  t))$. One thus sees that our leading order metric corresponds to the SdS metric, where {\it all} occurrences of $m$ and $\Lambda$ are replaced by their time-dependent counterparts $m(t), \Lambda(t)$. This way the metric is quasi-stationary in the strict sense: it corresponds exactly to a motion in the set of time-independent configurations.

Finally let us also point out that the functions $m(t)$ and $\Lambda(t)$ slightly differ between \eqref{fetheirs} and \eqref{feours}, i.e. between \cite{Gregory:2018ghc} and this paper. Although the ODEs determining these functions have the exact same form, they contain $\beta$ and $\gamma$, which are constant in \cite{Gregory:2018ghc} but functions of time here. More precisely, writing $\mathbf{u}(t)=(m(t),\Lambda(t))$, $\mathbf{c}=(\beta,\gamma)$ one has:
\begin{align}
	\mbox{\cite{Gregory:2018ghc}}\,:&\ \mathbf{EQ}\left(\mathbf{u}(t),\dot{ \mathbf{u}}(t);\mathbf{c}\right)=0\\
	\mbox{this paper}\,:&\ \mathbf{EQ}\left(\mathbf{u}(t),\dot{\mathbf{u}}(t);\mathbf{c}(\mathbf{u}(t))
	\right)=0	
\end{align} 
\color{black}
\subsection{The solution at next to leading order}\label{NLO}
Under the quasi-stationary expansion ansatz \eqref{slowtimexp}, the Einstein equations (\ref{feq1}-\ref{feq2}) at order $\calo(\epsilon)$ are
\begin{eqnarray}
\partial_r b_1&=&\frac{1}{\zeta_0}\frac{d}{d\tau}\frac{\zeta_0^2-1}{f_0}\, ,\label{b1eq}\\
\partial_r(r f_1)&=&\frac{\zeta^2_0-1}{f_0} \frac{r^2\dot\varphi^2}{4}\, ,\label{f1eq}
\end{eqnarray}
where $f_0, \zeta_0$ are given in (\ref{tdwarp}, \ref{zeta0}) and $\dot{\varphi}$ is determined in terms of $\alpha$ by the first of \eqref{ODEs}.
The equations (\ref{b1eq}, \ref{f1eq}) can be directly integrated and after some algebraic manipulations they lead to the simplified forms
\begin{align}
\hspace{-1 cm} & f_1(r,\tau)= \dot \alpha(\tau) \left(f_{10}(\tau)+\frac{f_{11}(\tau)}{r}+\frac{f_{12}(\tau)}{r} \log \frac{r}{l_1(\tau)}+\frac{f_{13}(\tau)}{r} \log\frac{r+\rbz(\tau)+\rcz(\tau)}{l_2(\tau)}-f_{14}(\tau)\,r^2 \right)\nonumber\, ,\\
\hspace{-1 cm} & b_1(r,\tau)= \dot \alpha(\tau) \left( b_{10}(\tau)+b_{11}(\tau)\log \frac{r}{l_3(\tau)}+b_{12}(\tau)\log\frac{r+\rbz(\tau)+\rcz(\tau)}{l_4(\tau)}+\frac{b_{13}(\tau)}{r+\rbz(\tau)+\rcz(\tau)}\right)\, ,\label{fb1}
\end{align}
where
\begin{align*}
\hspace{-2 cm}
f_{10}=- \frac{3^\frac{1}{4}2M\tan 3\alpha}{(\tan \a)^\frac{3}{2}}\,,\quad
f_{12}= -\frac{4M^2\sin 3\a}{\cos\a}\,,\quad
f_{13}= \frac{36 M^2\cos\a\sin 3\a}{\cos^2 3\a}\,,\quad
f_{14}=-\frac{(2+\csc^2 \a )^2 \sin 3 \a }{108\, (12)^{\frac{1}{4}}M\sqrt{ \sin 2\a}}\\
b_{11}=\frac{2\,(12)^\frac{1}{4} M(2 \sin \a + \sin 3 \a)}{(\sin  2 \a)^\frac{3}{2}}\,, \quad
b_{12}= \frac{3\,(12)^\frac{1}{4}M}{\cos 2 \a \sqrt{\sin 2 \a}}\,,\quad
b_{13}=- \frac{36 M^2 \sin 2 \a}{\cos^2 3\a}\,.
\end{align*}
The functions $f_{11}(\tau), b_{10}(\tau), l_a(\tau)$ are all left undetermined by the Einstein equations (\ref{b1eq}, \ref{f1eq}) but only two of them are independent since \eqref{fb1} is invariant under
\begin{equation*}
l_a\rightarrow \lambda_a l_a\, ,\quad f_{11}\rightarrow f_{11}+f_{12}\log\lambda_1+f_{13}\log\lambda_2\, ,\quad b_{10}\rightarrow b_{10}+b_{11}\log\lambda_3+b_{12}\log\lambda_4\, .
\end{equation*}
Out of the six functions $f_{11}(\tau), b_{10}(\tau), l_a(\tau)$ four can be chosen as preferred using the symmetry above, while the remaining two will be fixed by imposing the boundary conditions \eqref{bc2}.   At linear order in $\epsilon$ these read 
\begin{align}
\left.\left(2\partial_{\tau}f_1-b_1 r\frac{\dot\varphi^2}{2}-\frac{2\partial_\tau\partial_r f_0+\dot{\varphi}^2}{\partial_rf_0}f_1\right)\right|_{r=\rbz}&=0\, ,\label{bc1a}\\ \left.\left(2\partial_{\tau}f_1+b_1 r\frac{\dot\varphi^2}{2}-\frac{2\partial_\tau\partial_r f_0-\dot{\varphi}^2}{\partial_rf_0}f_1\right)\right|_{r=\rcz}&=0\label{bc1b}\, ,
\end{align}
where we used the leading order boundary conditions \eqref{bc0} and that
\begin{equation}
r_\mathrm{h}=r_{\mathrm{h}0}+\epsilon\, r_{\mathrm{h}1}+\calo(\epsilon^2)\,,\qquad r_{\mathrm{h}1}=-\left.\frac{f_1}{\partial_r f_0}\right|_{r=r_{\mathrm{h}0}}\,.
\end{equation}
valid for both the black hole horizon ($r_\mathrm{h}=\rb$) as well as the cosmic horizon ($r_\mathrm{h}=\rc)$. 
Let us shortly illustrate how (\ref{bc1a}, \ref{bc1b}) can be interpreted as two equations determining $f_{11} (\tau)$ and $b_{10}(\tau)$. To make this explicit, let us define
\begin{equation*}
f_1=\dot{\alpha}\left(\frac{f_{11}}{r}+\hat f_{1}\right)\, ,\qquad b_1=\dot{\alpha}\left(b_{10}+\hat b_1\right)\, .
\end{equation*}
Then (\ref{bc1a}, \ref{bc1b}) are equivalent to 
\begin{equation}\label{eqf11b10}
\partial_{\tau} f_{11} + \Gamma_+ f_{11} = \Delta_+ \, , \qquad b_{10} = \frac{4}{ \dot\varphi^2} \Big(\Gamma_{-} f_{11} - \Delta_- \Big)\, ,
\end{equation} 
with
\begin{equation*}
\Gamma_{+}=\frac{\rcz^2 \Gamma_{\mathrm{b}}+\rbz^2 \Gamma_{\mathrm{c}}}{\rbz^2+\rcz^2}\qquad \Delta_{+}=\frac{\rcz^2 \Delta_{\mathrm{b}}+\rbz^2 \Delta_{\mathrm{c}}}{\rbz^2+\rcz^2} \qquad \Gamma_{-}=\frac{\Gamma_{\mathrm{b}}-\Gamma_{\mathrm{c}}}{\rbz^2+\rcz^2}\qquad \Delta_{-}=\frac{\Delta_{\mathrm{b}}-\Delta_{\mathrm{c}}}{\rbz^2+\rcz^2}
\end{equation*}
where in turn
\begin{align*}
\Gamma_\mathrm{b}&=-\left.\frac{\dot{\alpha}\partial_\tau\left(\frac{\partial_r f_0}{\dot{\alpha}}\right)+\frac{\dot{\varphi}^2}{2}}{\partial_r f_0}\right|_{r=\rbz} \qquad \Delta_\mathrm{b}=-\rbz\left.\left(\partial_\tau \hat f_1+\Gamma_\mathrm{b}\hat f_1-\hat b_1 r\frac{\dot{\varphi}^2}{4}\right)\right|_{r=\rbz}\\ \Gamma_\mathrm{c}&=-\left.\frac{\dot{\alpha}\partial_\tau\left(\frac{\partial_r f_0}{\dot{\alpha}}\right)-\frac{\dot{\varphi}^2}{2}}{\partial_r f_0}\right|_{r=\rcz}
\qquad\Delta_\mathrm{c}=-\rcz\left.\left(\partial_\tau \hat f_1+\Gamma_\mathrm{c}\hat f_1+\hat b_1 r\frac{\dot{\varphi}^2}{4}\right)\right|_{r=\rcz}
\end{align*}
The equations \eqref{eqf11b10} are nothing but a simple ODE for $f_{11}$ and an algebraic expression for $b_{10}$. Furthermore, one can verify that $\Gamma_+$ has the form
\begin{equation}
\Gamma_+=-\partial_{\tau}\log I \qquad I=\frac{\sin^3\alpha\,\sqrt{\sin 2\a}}{\sqrt{2}\dot{\alpha}(1+2\sin^2\a)}
\end{equation}
and so by defining $f_{11}= I \tilde f_{11}$, the ODE on the left of \eqref{eqf11b10} reduces to $\partial_{\tau}\tilde f_{11}=\frac{\Delta_+}{I}$ which can be directly integrated. We have however not been able to simplify $\frac{\Delta_+}{I}$ to a form that is worth reproducing here, let alone can be explicitly integrated.
\\
\\
In conclusion, we find the metric \eqref{full lo + nlo metric} up to an error of $\mathcal{O}(\epsilon^2)$. As can easily be confirmed by looking at the functions $f_0, f_1, b_0, b_1$ and $\zeta_0$, this metric is well-behaved in between the two horizons. This confirms that the choice of slicing \eqref{scalarform} holds for a large patch of interest. We refer to subsection \ref{behaviour of the solutions} where the endpoints in time of our slicing condition are discussed. 
\\ Lastly, since $b_1 \neq 0$ (and thus also $a_1 \neq 0$), it follows that a more general timelike warp factor is required then in \cite{Gregory:2018ghc}.
\color{black}

\section{Conclusions and outlook}

In this paper, building upon the work of \cite{Gregory:2018ghc}, we constructed a spherical symmetric spacetime that solves the Einstein equations to second order in a quasi-stationary perturbation theory. The spacetime describes a dynamic Schwarzschild black hole coupled to a dynamic Sitter background through a time-dependent scalar field. This scalar field $\Phi(t)$ is minimally coupled to Einstein's gravity with boundary conditions imposing that it is purely ingoing on both horizons. As such, an evolution is introduced into the Schwarzschild-de Sitter (SdS) spacetime which, at leading order $\calo(\epsilon^0)$, amounts to quasi-stationary dynamics, i.e. where time dependence enters the metric only through the mass and cosmological constant parameters. 
\\This quasi-stationary approximation was introduced by assuming that all quantities depend on time only through the slow time variable $\tau = \epsilon t$ with $\epsilon \ll 1$ and then considering a series expansion in $\epsilon$. We found that the dynamics are such that the scalar field rolls down its potential increasing the mass of the black hole while the cosmic horizon moves further away in time.
\\ We also connect this quasi-stationary setup to the nonrelativistic $1/c$ expansion of GR \cite{Hartong2022}. This equivalence is to be expected since in both cases, time derivatives are suppressed which leads to small velocities. Interestingly, the solution up to the next-to-leading order is one for which no all-order closed-form solution is known. As such, the solution of this paper is also the first example of an application of the $1/c$ expansion beyond the exact solutions of GR and outside the Post-Newtonian regime.
\\
\\
Many possible extensions of our study remain. To start, it could be interesting to better understand the next-to-leading order metric as well as higher orders in $\epsilon$. Also, the connection with the nonrelativistic expansion can be elaborated.
\\
Once the Schwarzschild case is thoroughly understood, extensions include generalizing the results to more general black hole spacetimes by adding charge, angular momentum, different spacetime dimensions and higher-order curvature corrections in the action. Additionally, it would be interesting to see how the results get modified for a black hole in an Anti-de Sitter spacetime, in particular, due to possible applications in the AdS/CFT correspondence. 
\\ Finally, one of the biggest motivations for our study comes from applying quasi-stationary thermodynamic transformations on a black hole to understand its thermal properties better. As is well-known, Schwarzschild black holes have a negative heat capacity which is incompatible with equilibrium thermodynamics. It turns out that the heat capacity can still be calculated through a non-equilibrium method by using quasi-static evolution as worked out in this paper. This non-equilibrium derivation of the heat capacity will be discussed in the forthcoming paper \cite{Beyen23}.

\section*{Acknowledgements}
We thank N. Haider and C. Maes for their involvement in the initial stages of this research and T. Hertog and T. Van Riet for useful discussions and comments. DVdB was partially supported by the Bilim Akademisi through a BAGEP award.

\appendix
\section{Exact solution for reduced ansatz}\label{notimeap}
In this appendix, we show how, with the additional assumption\footnote{Remark that the weaker assumption $\partial_rA=0$ is equivalent to the choice $A=1$ since $A$ can then be put to one through a redefinition of the time coordinate, see \eqref{general time transformation}.} $A=1$ in the metric ansatz \eqref{metricform}, the equations (\ref{eq1}-\ref{eq3}) can be solved exactly in full generality. As mentioned in the main text, these equations are equivalent to the Euler-Lagrange equations of the theory \eqref{modellag}, in the case of {\it spherical symmetry}.

{For clarity, let us restate the above as an equivalent statement that is independent of the main text:
\\
\\
The 4d Einstein equations in the presence of energy-momentum of just a single canonical scalar field (with arbitrary potential) can be solved explicitly for all metrics of the form
\begin{equation}
	ds^2=-{F}dt^2+2H  dt dr+\frac{1-H^2}{F} dr^2+r^2 d\Omega^2\, , \label{simplemetric}
\end{equation}
with $F=F(t,r), H=H(t,r)$ arbitrary functions of $t$ and $r$, and one works in a gauge where $\Phi=\Phi(t)$.} \color{black} As shown \color{black} in this appendix, the only {\it dynamic}, i.e. non-stationary, metric among such solutions is an FLRW metric (without a black hole). {The metric being dynamic is equivalent to the assumption $\dot \Phi\neq 0$, which we'll make below. 
\\
\\
As is well known, the only stationary solutions of the form \eqref{simplemetric} are those in the SdS family (i.e., in this case, a black hole can be present). Let us stress that the absence of black holes in the dynamic case is completely due to the restriction to $A=1$. Indeed, as is shown in the main text, dynamic black holes \textit{can} appear when $A$ is allowed to become a non-trivial function. This is all perfectly consistent due to the observation that for the (perturbative) solutions of the main text $A=1+A_1(r,t)\epsilon+\ldots$\. In the limit where one turns off the time-dependence, or makes it asymptotically slow, i.e. $\epsilon\rightarrow 0$, one sees that $A\rightarrow 1$, recovering the (stationary) SdS family. Thus, the dynamic black holes simply become an SdS black hole as one dials down the dynamics. Or in reverse, as one slowly introduces time dependence into an SdS black hole the warp-factor $A$ will slightly start deviating from $1$ and the metric is taken outside of the ansatz \eqref{simplemetric} into the more general class \eqref{metricform}. }
\\
\\
{In this appendix, we take equations (\ref{eq1}-\ref{eq3}) as our starting point.}
The assumption $A=1$ immediately implies through \eqref{eq1} that the metric component $g_{rr}$ will be time-independent, i.e.
\begin{equation}
g_{rr} = \frac{1-H^2}{F}=q(r)\, ,\label{grreq}
\end{equation}
for some arbitrary function $q$. In turn, equation \eqref{eq3} can then be integrated to give
\begin{equation}
F = 1 - \frac{2 m(t)}{r} - \frac{\L(t)}{3}r^2 - \frac{\dot \Phi^2}{4r } p(r) \quad \text{with} \quad p(r) = \int q(r) r^2 dr\, .\label{Fexact}
\end{equation}
Here $m(t)$ is an arbitrary function and $\Lambda(t)=\frac{V(\Phi(t))}{2}$. Note that the time-dependent integration constant is set to be $ -2m(t)$ for convenience. In addition to \eqref{Fexact} one is then left with \eqref{eq2} and \eqref{grreq} which take the form 
\begin{eqnarray}
H^2 &=& 1 - \frac{p'}{r^2} F\, , \label{eq12}\\
\dot F &=& - \frac{r H \dot \Phi^2 }{2 } \, . \label{eq22}
\end{eqnarray}
Combining the time derivative of \eqref{eq12} with \eqref{eq22} gives 
\begin{equation*}
\dot H = \pm \frac{p'}{4 r} \dot\Phi^2 \, ,
\end{equation*}
which can be integrated to
\begin{equation}
H = H_0(r)  \pm \frac{p'}{4 r} \o\,,\qquad \o = \int^t \dot\Phi^2 d\tilde t\, . \label{hsol}
\end{equation}
Finally inserting \eqref{hsol} and \eqref{Fexact} into \eqref{eq12} one finds the equation
\begin{eqnarray}
\frac{p'}{r^2} \left(1 - \frac{2 m}{r} - \frac{\L}{3}r^2 - \frac{ p}{4r } \dot \o \right) &=& 1-\left(H_0 \pm \frac{p'}{4 r} \o\right)^2\, .  \label{finaleq}
\end{eqnarray}
We will assume $\dot \o \neq 0$ and $p'\neq 0$ for now and reconsider the other cases separately at the end of this appendix. To proceed, we solve equation \eqref{finaleq} for $p$, which can then be re-inserted into \eqref{eq12} and \eqref{Fexact} to obtain the full solution. 
\\First we make a technical simplification\footnote{One could repeat the argument here by keeping $t$ as the variable and instead of taking repeated $\o$ derivatives, one would simply take $t$-derivatives. Through the chain rule, these derivatives are related as $\frac{d}{d\o} = \frac{1}{\dot \o} \frac{d}{dt}$.} by using $\omega$ as a variable instead of $t$. I.e. we formally invert $\omega(t)$ to $t(\omega)$ and then think of all unknown functions of $t$ as functions of $\omega$ instead:
\begin{equation*}
m(\omega) = m(t(\o)) \qquad \L(\omega) = \L (t(\o)) \qquad \dot \o(\omega) = \dot \o (t(\o))\, 
\end{equation*}
We then take the $\o$-derivative of \eqref{finaleq} twice to find
\begin{eqnarray}
\frac{2}{r }\frac{d m}{d\o } + \frac{ r^2}{3  }\frac{d \L}{d\o} +\frac{p}{4r }\frac{d \dot \o}{d\o } &=&   \frac{r}{2}\left(\pm H_0  + \frac{p'}{4r} \o \right)\, , \label{1der}\\
\frac{2}{r }\frac{d^2 m}{d\o^2} + \frac{ r^2}{3  }\frac{d^2 \L}{d\o^2} +\frac{p}{4r }\frac{d^2 \dot \o}{d\o^2} &=& \frac{p'}{8}\, .\label{2der}
\end{eqnarray}
Taking a final $\o$-derivative of \eqref{2der}, leads to an algebraic equation for $p$ whose unique solution is\footnote{Here we assume $\frac{d^3\dot{\omega}}{d\omega^3}\neq 0$, the case where $\frac{d^3\dot{\omega}}{d\omega^3}=0$ is discussed separately at the end. We thank R. Gregory for alerting us to that special case.}:
\begin{eqnarray}
p (r) =-  \frac{8b_0}{c_0} - \frac{4}{3c_0}  r^3 \quad \text{where} \quad b_0= \frac{\frac{d^3 m}{d \o^3}}{\frac{d^3  \L}{d\o^3}}  \quad c_0= \frac{\frac{d^3 \dot \o}{d\o^3}}{ \frac{d^3 \L}{d \o^3}} \, . \label{psolution}
\end{eqnarray}
Since $p=p(r)$ is a function of $r$ only, $b_0$ and $c_0$ have to be time-independent constants\footnote{Remark that $b_0$ and $c_0$ must also be independent of $r$ since $m$ and $\Lambda$ are functions of $\omega$, or equivalently $t$, only.}. This result then allows us to immediately integrate the definitions on the right in \eqref{psolution} to
\begin{eqnarray}
m = b_0 \L + b_1 \frac{\o^2}{2}+b_2 \o + b_3 \qquad \dot \o = c_0 \L + c_1 \frac{\o^2}{2}+ c_2 \o + c_3 \, . \label{mLexact}
\end{eqnarray}
Here all $b_i, c_i$ are numeric constants, most of them are fixed by inserting \eqref{psolution} back into \eqref{finaleq}, \eqref{1der} and \eqref{2der}. One finds that all of them can be expressed in terms of $b_0$ and $c_2$ as
\begin{align*}
b_1 &= -\frac{3b_0}{8}\, ,&  b_2 &=-\frac{b_0 c_2}{4} \, , &b_3 &=-\frac{b_0c_2^2}{12}\, , \\
c_0 &= -4 \, ,&  c_1 &=\frac{3}{2} \, , & c_3 &=\frac{c_2^2}{3}\, ,
\end{align*}
and furthermore $H_0(r) = \pm \frac{r c_2}{6}$. Now that the dust has settled, one finds that the functions appearing directly in the metric are
\begin{equation}
F = 1- \left(\frac{c_2}{6} + \frac{\o}{4} \right)^2 r^2 \, , \qquad H =  \pm \left( \frac{c_2}{6}  + \frac{\o}{4 }  \right)r\, . \label{finalmetricfunctions}
\end{equation}
Remark that we can set $c_2=0$ without loss of generality since $\omega=\int^t \dot\Phi^2 d\tilde t$ was only determined up to an overall constant in any case.
\\ Lastly, performing a change of coordinates $r = a(t) \rho$ and requiring that this transformation makes the metric diagonal results in the equation
\begin{equation}
\dot a(t) = -\frac{1}{4}a(t)\o(t)\, .\label{scalefactor}
\end{equation}
The final metric is then
\begin{equation}
ds^2 = -dt^2 + a(t)^2(d\r^2 + \r^2 d\O_2^2) \quad \text{where} \quad a(t) = a_0 \, e^{- \frac{1}{4}\int^t\int^{t'} \dot\Phi^2(t'') \ dt'' \ dt'}\, , \label{flrwmetric}
\end{equation}
which is the standard metric for a FLRW universe. Furthermore one verifies that the equation on the right of \eqref{mLexact} is equivalent to the Friedmann acceleration equation for $a$, with $\r = p=\frac{\dot \Phi^2}{4}$. 
\\
\\
Finally, to complete the discussion, we should mention the cases where the assumptions made above, $\dot \o \neq 0, \frac{d^3\dot{\omega}}{d\omega^3}\neq 0$ and $p'\neq 0$, are not satisfied. 
\\
In case $\dot \o = \dot \Phi^2 = 0$, the scalar field is a constant $\Phi=\Phi_\star$ and one finds that the unique\footnote{Remark that $H$ remains an arbitrary function of $r$, that can be removed by a coordinate transformation to put the metric in the manifestly static form \eqref{SdSsol}.} solution to the equations is the SdS metric with $\Lambda=\frac{V(\Phi_\star)}{2}$ discussed in section \ref{SdSsection}.
\\
For $p' = 0$, it follows immediately from \eqref{eq12} that $H=\pm 1$ and combining (\ref{Fexact},  \ref{eq22}) then imply $\dot m=0$, $\dot{\Lambda}=0$ and $\dot{\Phi}=0$. So also in this case we are back to the satic SdS solution. To put it into the form \eqref{SdSsol}, one should now perform the coordinate transformation $t=t'+\int^r \frac{dr'}{F}$.
\\
\\
Finally, we should check the case $\frac{d^3\dot{\omega}}{d\omega^3}= 0$, which turns out to be somewhat subtle. The analysis up to and including \eqref{2der} remains unchanged, but upon taking a further $\omega$ derivative of \eqref{2der} one now finds that also the third derivative of $\Lambda(\omega)$ and $m(\omega)$ has to vanish. One can thus write
\begin{equation}
\dot{\omega}=a_0+a_1 \omega+\frac{a_2}{2}\omega^2\qquad \Lambda=b_0+b_1 \omega+\frac{b_2}{2}\omega^2\qquad m=c_0+c_1 \omega+\frac{c_2}{2}\omega^2 \label{3dersol}
\end{equation}
with $a_i, b_i$ and $c_i$ constants. All $r$ and $t$ (or $\omega$) dependence is now set.  By reinserting \eqref{3dersol} into (\ref{1der}, \ref{2der}) one finds equations for $p$ and $H_0$ that can be explicitly solved. This then fixes the $r$ dependence of \eqref{finaleq} completely and one obtains a number of constraints by the requirement that each coefficient in an expansion in $r$ has to vanish separately. 
\\ The analysis of these solutions, although straightforward, is somewhat involved and cumbersome because the type of $r$ dependence -- which powers that appear as well as possible logarithmic terms -- depends on the value of the constant $a_2$. Without providing details let us simply list the results. For all values of $a_2$, there is a solution where $F=1-H^2$, $H=r(q_1+q_2 \omega)$ with $q_1$ and $q_2$ constants. Such a metric can always be put into FLRW form by the same transformations as above. For generic values of $a_2$, this FLRW solution is the only one, but for a few specific values of $a_2$ an additional solution exists as well. When $a_2=-\frac{1}{2}$, the extra solution has $\dot{\omega}=\dot{\Phi}^2<0$ and is hence unphysical, while for $a_2=\frac{1}{2}$ the additional solution that appears is nothing but dS. The only remaining case happens when $a_2=\frac{5}{2}$, which is less trivial to analyze.
Here, one finds that $\dot{\omega}=\frac{5}{4}\left(\omega+\frac{2}{5}a_1\right)^2$, which can be integrated to $\omega=-\frac{2}{5}\left(a_1+\frac{2}{t}\right)$. Via $\Lambda=\frac{V}{2}$ and $\dot{\omega}=\dot{\Phi}^2$, which can be integrated for $\Phi(t)$, it follows that
\begin{equation}
V=-\frac{4}{25} e^{-\sqrt{5} \Phi} \left(1 +\lambda e^{\frac{\sqrt{5}}{2}\Phi}\right)\label{specialpot}
\end{equation}
where $\lambda=10b_1+a_1$ is an arbitrary constant. As such, this additional solution only exists for potentials of the particular form \eqref{specialpot}. Working out $F$ and $H$, one finds the metric
\begin{eqnarray}
ds^2&=&-\left(1+\frac{r^2 \left(\lambda ^2 r^2+150 \lambda  t-225\right)}{5625 t^2}\right)dt^2+\frac{2 r\left(\lambda ^2 r^2+75 \lambda  t-225\right)}{1125 t}dtdr\nonumber\\
&&+\left(1-\frac{\lambda ^2 r^2}{225}\right)dr^2+r^2d\Omega^2
\end{eqnarray}
We have not found a coordinate transformation that would simplify this metric further. One can check that the metric above is of Petrov type D and as such it could be interesting to see if it can be identified among the solutions listed in e.g. section 33.3 of \cite{stephani2009exact}.

\section{Horizons and boundary conditions}\label{bcapp}
Here we present some technical details on the argument in section \ref{bcsec} that led to the boundary conditions \eqref{bc}. We show that the boundaries $r=\rb$ and $r=\rc$ are apparent horizons and that \eqref{bc} guarantees that disturbances of the scalar field $\Phi$ can only propagate into these horizons, not emerge from them.
\\
\\
We will consider metrics of the form \eqref{metricform}, with the additional assumption that $F(r,t)$ has exactly two positive zeros, which correspond to $r=\rb(t)$ and $r=\rc(t)$, chosen such that $0\leq \rb(t)\leq \rc(t)$. One should remark that for generic time dependence the hypersurfaces $r=\rb(t)$ and $r=\rc(t)$ are not null and so these surfaces are not event horizons. As we will discuss below they both correspond to a more general notion of horizon, that of {\it apparent horizon}  as defined in e.g. \cite{Faraoni2015}. We will restrict attention to the region $\calr$ defined by $\rb< r< \rc$ and also assume that $F>0$, $A>0$ and $-1<H<1$ in this region, which is the case in the solutions discussed in this paper. 
\\
Consider then the following vector fields, which are all null with respect to \eqref{metricform}:
\begin{equation}
l_\pm=(1\pm H)\partial_t\pm AF \partial_r\,,\qquad n_{\pm}=\frac{1}{F}l_\pm=\frac{1\pm H}{F}\partial_t\pm A \partial_r\label{ln}\,.
\end{equation}
In $\calr$, all of them are future-oriented, with $l_+$ and $n_+$ pointing radially outward while $l_-$ and $n_-$ are directed radially inward.  Note that the $n_\pm$ are introduced as auxiliary fields\footnote{Remark that $l_+^\m l^-_\m=-2FA$ so that they are only linearly independent away from $F=0$. By introducing the $n_{\pm}$, we have pairs that remain linearly independent even when $F=0$, which is important when checking the conditions determining the type of horizon below. Although one might fear that $n_\pm$ become ill-defined on $F=0$ where they naively blow up, this is only a coordinate artefact, since $l_\pm^\mu n^\mp_\m$ remains finite there.} so that each of the pairs $(l_\pm, n_\mp)$ span the 2-dimensional subspace transverse to the sphere, i.e. $l_{\pm}^\m n^\mp_\mu=-2A<0$. 
\\ 

\paragraph{Horizons}  Following e.g. \cite{poisson_2004, black_hole_physics_book, Faraoni2015}, we can look for the presence of horizons by investigating the expansions of congruences of null geodesics. Computing these for the pair $(l_+, n_-)$ one finds that
\begin{eqnarray}
\theta_{l_+}&=&\left(g^{\m\n}-\frac{n_-^\m l_+^\n+l_+^\m n_-}{l_+^\r n^-_\r}\right)\nabla_\m l^+_\n=\frac{2 A F}{r}\, ,\label{leq}\\
\theta_{n_-}&=&\left(g^{\m\n}-\frac{l_+^\m n_-^\n+n_-^\m l_+^\n}{l_+^\r n^-_\r}\right)\nabla_\m n^-_\n=- \frac{2 A}{r}\, .\label{neq}
\end{eqnarray}
\\
Equations (\ref{leq}, \ref{neq}) tell us that in the region $\calr$, the congruence of radially outgoing null geodesics tangent to $l_+$ is expanding, while that tangent to $n_-$ -- which is radially ingoing -- is contracting. This is similar to the situation in flat space, where gravity is absent. The key point is however that, at $r=\rb$ and $r=\rc$, this behaviour changes since the expansion \eqref{leq} vanishes there so that the outgoing geodesics are now on the brink of starting to contract, while simultaneously the ingoing geodesics keep contracting. This behaviour is what defines the notion of {\it apparent horizon}, see e.g. \cite{Faraoni2015}. Stated precisely:
\begin{equation} 
r=\rb\mbox{ or }r=\rc\quad \Rightarrow \qquad \theta_{l_+}=0\,,\quad \theta_{n_-}<0\label{hcond}\\
\end{equation}
identifies both surfaces $r=\rb$ and $r=\rc$ as apparent horizons. Although these are not event horizons, there are good physical arguments, see again e.g. \cite{Faraoni2015}, to give $r=\rb$ the interpretation of a black hole horizon and $r=\rc$ that of a cosmological horizon. In the special case where $\rb$ and $\rc$ are time independent all the technical results above are still valid, and in this case, these apparent horizons become actual event horizons, see e.g. the Penrose diagram in Figure \ref{Penrose}. 

\paragraph{Boundary conditions} From the point of view of the region $\calr$ between the horizons, $l_-$ is ingoing on the black hole horizon $r=\rb$ while $l_+$ is ingoing on the cosmological horizon $r=\rc$. As such, they point from $\calr$ into the spheres at its boundary. A scalar field $\Phi$ can thus be said to be purely ingoing\footnote{To compare: in 2d Minkowski space $ds^2_{\mathrm{2d}}=dx^+dx^-$, $x^\pm=t\pm r$, a scalar field $\Phi(x^+,x^-)=f(x^-)$ is purely rightmoving, and thus ingoing on a boundary $r=r_0$ from the point of view of the region $r<r_0$. This condition equivalently corresponds to demanding that $\partial_{x_+}\Phi=0$. Similarly, a scalar field that satisfies $\partial_{x_-}\Phi=0$ is purely left moving and hence ingoing on such a boundary from the point of view of the region $r>r_0$.} on the horizons iff
\begin{equation}
\left.l_-\Phi\right|_{r=\rb}=0\, ,\qquad \left.l_+\Phi\right|_{r=\rc}=0 \, .\label{vecbcs}
\end{equation}
Since we have defined the coordinate $t$ such that $\Phi=\Phi(t)$, i.e. $\partial_r\Phi=0$, the conditions \eqref{vecbcs} are equivalent\footnote{Naively one might worry that \eqref{bc} imply that $l_-$ (respectively $l_+$) vanishes on $\rb$ (respectively $\rc$) since $F$ vanishes on the horizon. This is however not the case, since $F\approx F'(r-\rb)$  near the horizon and thus $F\partial_r\approx F' (r-\rb)\partial_r= F'\partial_{\rho}$ via $\rho=\log(r-\rb)$ and so in well defined local coordinates one sees that $l_-$ (and similarly $l_+$ ) remains non-zero on the horizon.} to \eqref{bc}.

\section{Toy friction model}\label{toyap}
In this appendix, we apply the quasi-stationary expansion to a simple mechanical system to provide some intuition to its application to GR with a scalar field.

Consider then a simple model of a 1-dimensional particle with friction:
\begin{equation*}
L=e^{\beta t}\left(\frac{m}{2}\dot x^2-V(x)\right)
\end{equation*}
Its equation of motion is
\begin{equation}\label{1d eq t}
m\ddot x+\beta m \dot x=-V'(x)
\end{equation}
which has the same structure of time derivatives as the scalar equation \eqref{scaleq}.
Assuming that $x=x(\epsilon t)$, writing $\tau=\epsilon t$ and reinterpreting in what follows the symbol $\dot{}$ as the $\tau$ derivative, \eqref{1d eq t} becomes
\begin{equation}\label{1d eq tau}
\epsilon^2 	m\ddot x+\epsilon\beta m \dot x=-V'(x)
\end{equation}
Similar to the setup discussed in section \ref{setup}, we assume the potential has the form $V(x)=U(y)$ with $y=\epsilon x$ such that \eqref{1d eq tau} reduces to
\begin{equation}\label{eq with y}
\epsilon 	m\ddot y+\beta m \dot y=-U'(y)
\end{equation}
So in this particular way of setting up the expansion, we have a slow time expansion where a non-constant potential remains present at leading order. Equation \eqref{eq with y} can directly be obtained from the Lagrangian\footnote{Note that $Ldt=\epsilon^{-1}L d\tau$, so, strictly speaking, we should multiply the Lagrangian with $\epsilon^{-1}$ as well, but since this has no effect at this level of the discussion we'll ignore that.}
\begin{equation}
L=e^{\frac{\b}{\epsilon}t}\left(\frac{m\epsilon}{2}\dot y^2-U(y)\right)
\end{equation}
\paragraph*{Example} For the potential $U(y)=\frac{k y^2}{2}$ and defining $\alpha=\frac{k}{m\beta^2}$, the exact solution takes the form
\begin{eqnarray}
y(\tau,\epsilon)&=&a(\epsilon) e^{- \frac{1-\sqrt{1-4\alpha \epsilon}}{2\epsilon}\beta\tau}+b(\epsilon) e^{-\frac{\beta\tau}{\epsilon}} e^{\frac{1-\sqrt{1-4\alpha \epsilon}}{2\epsilon}\beta\tau} \label{solution y} \\
&=&\sum_{n=0}^\infty y_n(\tau) \epsilon^{n}+e^{-\frac{\beta\tau}{\epsilon}}\sum_{n=0}^\infty \tilde y_n(\tau) \epsilon^{n} \label{trans}
\end{eqnarray}
With the quasistationary setup of this problem in mind, it is mostly interesting to observe the transseries form of the solution. The second series in \eqref{trans} will, in a perturbative approach, only be reproduced upon careful resummation/resurgence of the first series \cite{resurgence}. Furthermore, the non-analytic contribution $\sim e^{-\frac{\beta\tau}{\epsilon}}$ 
also shows that the small $\epsilon$ approximation only works for $\beta>0$.
\\The fact one loses the integration constant $b$ in a perturbative approach can now be seen both from the solution, where it multiplies the non-analytic term, as well as from \eqref{eq with y}, where at leading order, the equation goes from a second order ODE into a first-order one. A similar feature is present in our GR setup. At leading and subleading order the perturbative solution takes the form
\begin{equation*}
y(\tau,\epsilon)=e^{-\alpha \beta\tau }\left(a_0+(a_1-\alpha^2 \beta \tau\, a_0)\epsilon+\calo(\epsilon^2)\right) 
\end{equation*} 
which upon translating back to $x$ gives
\begin{equation*}
x(t,\epsilon)=\frac{e^{-\alpha \beta \epsilon t}}{\epsilon}\left(a_0+(a_1-\alpha^2 \beta \epsilon t\, a_0)\epsilon+\calo(\epsilon^2)\right) 
\end{equation*}

This example provides a concrete and fully solvable realization of the quasi-stationary expansion, showing some of its merits as well as limitations.

\bibliographystyle{utphys}
\bibliography{SdS_lit}

\end{document}